\documentclass{aastex62}
\usepackage{tabularx}
\usepackage{graphicx, nicefrac}
\hypersetup{linkcolor=blue,citecolor=blue,filecolor=blue,urlcolor=blue}

\graphicspath{{./}{figures/}}

\begin{document}

\title{Five new hot-Jupiter transits investigated with \textsl{Swift}-UVOT}

\correspondingauthor{L.~Corrales}
\email{liac@umich.edu}

\author[0000-0002-5466-3817]{Lia Corrales}
\affil{University of Michigan, Dept. of Astronomy \\
1085 S University Ave \\
Ann Arbor, MI 48109, USA}

\author[0000-0002-8951-6139]{Sasikrishna Ravi}
\affil{University of Michigan, Dept. of Astronomy \\
1085 S University Ave \\
Ann Arbor, MI 48109, USA}

\author[0000-0002-3641-6636]{George W. King}
\affil{University of Michigan, Dept. of Astronomy \\
1085 S University Ave \\
Ann Arbor, MI 48109, USA}

\author{Erin May}
\affil{Johns Hopkins University, Applied Physics Laboratory \\
11100 Johns Hopkins Rd \\
Laurel, MD 20723-6099, USA}

\author{Emily Rauscher}
\affil{University of Michigan, Dept. of Astronomy \\
1085 S University Ave \\
Ann Arbor, MI 48109, USA}

\author{Mark Reynolds}
\affil{University of Michigan, Dept. of Astronomy \\
1085 S University Ave \\
Ann Arbor, MI 48109, USA}

\begin{abstract}

Short wavelength exoplanet transit measurements have been used to probe mass-loss in exoplanet atmospheres. We present the \textsl{Swift}-UVOT transit lightcurves for five hot Jupiters orbiting UV-bright F-type stars: XO-3, KELT-3, WASP-3, WASP-62, and HAT-P-6. We report one positive transit detection of XO-3b and one marginal detection of KELT-3b. We place upper limits on the remaining three transit depths.  
The planetary radii derived from the NUV transit depths of both potential detections are $50-100\%$ larger than their optical radius measurements. We examine the ratio $R_{\rm NUV}/R_{\rm opt}$ for trends as a function of estimated mass-loss rate, which we derive from X-ray luminosity obtained from the \textsl{Swift}-XRT, or XMM-Newton in the case of WASP-62. We find no correlation between the energy-limited photoevaporative mass-loss rate and the $R_{\rm NUV}/R_{\rm opt}$ ratio. We also search for trends based on the equilibrium temperature of the hot Jupiters. We find a possible indication of a transition in the $R_{\rm NUV}/R_{\rm opt}$ ratio around $T_{\rm eq} = 1700$~K, analogous to the trends found for NIR water features in transmission spectra. This might be explained by the formation of extended cloud decks with silicate particles $\leq 1~\mu{\rm m}$. We demonstrate that the \textsl{Swift}-UVOT filters could be sensitive to absorption from aerosols in exoplanet atmospheres.
\end{abstract}


    \section{Introduction} \label{sec:intro}

More than than 75\% of the 4,300 confirmed exoplanets today have been detected via the transit method.\footnote{NASA Exoplanet Archive (\href{https://exoplanetarchive.ipac.caltech.edu/}{https://exoplanetarchive.ipac.caltech.edu})}
By combining transit measurements from multiple wavelengths, one can measure the apparent radius of a planet as a function of wavelength, giving a rough description of the planetary atmosphere's absorbant properties. The resulting transmission spectrum can be fit for temperature and pressure profiles, chemical composition, and the presence of key life ingredients such as water \citep[e.g.,][]{Charbonneau2002,Madhusudhan2019}. 
Hot Jupiters are the ideal first targets for transmission spectroscopy because of their deep transits and short orbital periods. 
In this work, we examine what short wavelength ($< 300$~nm) transits contribute to the science of hot Jupiter atmospheres. In some cases, short wavelength transits are much deeper than observed in the optical. Such observations have provided evidence for planetary exospheres that are extended beyond the Roche lobe. The strongest of these are the Ly$\alpha$ transit observations of HD~209458b \citep{Vidal-Madjar2003, Ehrenreich2008}, HD~189733b \citep{Lecavalier2010, Lecavelier2012}, GJ~436b \citep{Kulow2014,Ehrenreich2015}, GJ~3470b \citep{Bourrier2018}, and recently HD~6433c \citep{Zhang2021}. In the NUV, FeII and MgII absorption from an extended atmosphere have been detected for WASP-121b \citep{Salz2019, Sing2019} and HD~209458b \citep{VidalMadjar2013, Cubillos2020}.

\citet[][hereafter S19]{Salz2019} demonstrates the capability of the Neil Gehrels Swift Observatory UVOT instrument to obtain transit depths with 0.5\% precision, measuring a $2.1 \pm 0.3\%$ NUV transit depth for the ultrahot Jupiter WASP-121b. Close to the optical, the apparent planet radius is expected to grow with shorter wavelengths due to Rayleigh scattering from high altitude molecules \citep[e.g.,][]{Sing2011}. However, S19 found that the NUV transit of the ultra-hot Jupiter WASP-121b is deeper than could be explained by this phenomenon. 
Recent work by \citet{Lothringer2020} demonstrates that, for irradiated planets with equilibrium temperatures $> 1600$~K, the NUV band is dominated by absorption from Fe and Mg, which exceeds the effect of Rayleigh scattering.
Indeed, S19 concluded that FeII was the dominant source of absorption in this system, which has since been identified in higher resolution transit measurements of WASP-121b \citep{Sing2019}. S19 also concluded that, because the broad-band NUV transit radius was larger than that observed for WASP-80b \citep{King2018}, which has an estimated mass loss rates that is 1000 times lower than that for WASP-121b, deep NUV transits might be positively correlated with mass-loss rates.

We also wish to examine how aerosols in exoplanet atmospheres might be probed by short wavelength spectroscopy. 
Hot Jupiters, despite their relatively high atmospheric temperatures, show signatures of obscuring aerosols \citep{Sing2016}. 
Aerosol candidates fall into two main categories: hazes, which are complex molecules typically formed through chemical reactions catalyzed by UV photons, and clouds, which condense out of the atmosphere. 
If UV photons are used up in chemical reactions, it is natural to expect that the UV absorptivity of a haze-obscured atmosphere is particularly high, causing a deeper UV transit. As an example, the photodissociation cross-section for common molecules (e.g., H$_2$O, NH$_3$, and CO$_2$) peak in the 50-300~nm wavelength range \citep[e.g.,][]{SPARCSwp} and lab produced aerosols such as CHON-rich tholins have large absorption opacities in the 100-250~nm range \citep{Gavilan2018}. 
Models of cloud formation in exoplanet atmospheres predict a range of cloud particle sizes from 1~nm to 1~cm, with smaller particles extending above the main cloud deck by many orders of magnitude in pressure \citep{Helling2016, Helling2019, Helling2020, Powell2018, Powell2019}. 
An extended layer of cloud particles $\lesssim 0.1~\mu{\rm m}$ in size can be particularly efficient at attenuating UV light, as seen in the interstellar medium (ISM) \citep[e.g.,][]{MRN1977}. 
Thus NUV transit measurements of hot Jupiters may provide some insight into clouds and hazes.

In this work, we use {\sl Swift}-UVOT to extend short wavelength transit observations to five additional hot Jupiters orbiting UV bright stars: WASP-3b \citep{Pollaco2008}, WASP-62b \citep{Hellier2012}, XO-3b \citep{Winn2008}, KELT-3b \citep{Pepper2013}, and HAT-P-6b \citep{Noyes2008}. 
In Section~\ref{sec:analysis}, we describe our \textsl{Swift}-UVOT dataset and analysis techniques, and demonstrate that they are sufficient to reproduce the work of S19. We also search for the X-ray emission from the exoplanet host stars using the simultaneous \textsl{Swift}-XRT observations. In Section~\ref{sec:results}, we discuss the results of the transit fitting for each target. In Section~\ref{sec:RNUVcompare}, we examine trends in the apparent NUV radius of each planet in relation to the X-ray insolation from the host star, estimated mass-loss rates, and equilibrium temperature.  We summarize our conclusions in Section~\ref{sec:conclude}. 

    \section{Data Analysis} \label{sec:analysis}

Targets were selected by identifying UV bright stars with hot Jupiter transits on the order of $1\%$. \textsl{Swift} monitored 2-3 transits for each of our targets, providing short $\sim 1-2$~ks snapshots over the course of a 10-15~ks window centered on each transit. We refer to the set of observations covering a single transit as a ``visit''.  
The transit observations were planned using the ephemeris calculator supplied by the NASA Exoplanets Archive (NEA).
Estimates for the observational UV magnitudes were obtained from the MAST GALEX source catalog, and each target was anticipated to yield $30-300$ counts per second in the \textsl{Swift}-UVOT \textsl{uvm2} filter. 
This filter was selected because it has the least sensitivity to optical light. 
The majority of the UVOT observations were taken in photon-counting (event) mode, and we limit our analysis to event mode data because it allows for customized binning of the host star lightcurve. 

Table~\ref{tab:vislist} lists the basic properties of our sample including host star characteristics, visit dates, and number of field stars used for calibration (N$_{\rm ref}$, described in \S\ref{sec:reduction}). The properties of the \textsl{Swift}-UVOT datasets are also summarized, including the total event mode exposure time, the duration of each transit, and the percentage of the transit covered by the observations. 
As a proof-of-concept, we also applied our analysis techniques to WASP-121b, originally reported by S19. %; the properties of this \textsl{Swift} dataset are also described throughout this work.

\begin{table}[htp]
    \centering
    \caption{\textsl{Swift}-UVOT datasets used for UV transit measurements.}
    \label{tab:vislist}
    \begin{tabular}{l c c l c c c c c c} 
        \hline
        & \textbf{Spectral} & \textbf{GALEX} & \textbf{Visit date} &  \textbf{Exposure} & \textbf{T}$_{14}$ & \textbf{transit} & \textbf{N} & \textbf{Rate} & \textbf{Binsize} \\
        & \textbf{Type}  & NUV & (YY-MM-DD)  & [hours] & [hours] & \textbf{covered} & \textbf{ref} & [ct/s] & [s]\\ 
        \hline \hline
        \textbf{XO-3} & F5V & 14.0 & 19-07-14, 19-10-05 & 3.364 & 2.989 & 55\% & 3 & 48 & 65 \\
        \textbf{KELT-3} & F7V & 14.1 & 19-05-31, 19-06-11, 20-02-14 %, 20-02-24 
        & 5.008 & 3.153 & 70\% & 3 & 30 & 110 \\
        \textbf{WASP-3} & F7-8V & 14.8 & 19-07-24, 19-08-15, 19-08-26 & 3.400 & 2.772 & 50\% & 5 & 20 & 160 \\
        \textbf{WASP-62} & F9V & 14.9 & 19-05-13, 19-05-22, 19-06-04 & 3.006 & 3.811 & 25\% & 4 & 17 & 190 \\
        \textbf{HAT-P-6} & F8V & 14.2 & 19-09-13, 19-10-02 & 3.286 & 3.506 & 50\% & 1 & 33 & 91 \\
        \hline
        \textbf{WASP-121} & F6V & 14.8 & 17-08-04, 17-09-03, 17-09-05 & 6.203 & 2.886 & 100\% & 7 & 20 & 160 \\
        \hline
        \hline
        \multicolumn{9}{l}{\textbf{Note.} \text{T}$_{14}$ is the transit duration, i.e., the time between first and last contact} \\
        \hline
    \end{tabular}
\end{table}

\subsection{Data Reduction}
\label{sec:reduction}

All \textsl{Swift} data reduction was performed with HEAsoft v6.28 and \textsl{Swift}-UVOT CALDB files released on October 26, 2020.  
With the aim of achieving $<2\%$ photometric precision, we used the event files to select time intervals to have a minimum of 3000 counts per bin from target star. As a result, the light curve bin size is different for each source, depending on the count rate of each host star. Table~\ref{tab:vislist} lists the average count rate for each target along with the median bin size used. We used the \texttt{barycorr} tool to convert the time bin values to barycentric time. Then we used \texttt{uvotscreen} and \texttt{uvotimage} to create images for each time bin.

To align the images, we ran \texttt{uvotdetect} on a single long exposure observation to create a point source catalog for the field. This catalog was fed into \texttt{uvotskycorr} to perform aspect corrections before running \texttt{uvotimsum} on image files with multiple image extensions. For targets with only a few stars within the \textsl{Swift}-UVOT field-of-view, aspect corrections were not always successful. Summing the image extensions then led to extended point source images. %and, occasionally, double images of the target stars. --- As I understand it, at least the double images disappeared after we employed our re-alignment.
We correct as much as possible for this effect by performing photometry on reference stars, described below, which will also suffer from aspect correction artifacts. In general, poor aspect corrections reduced the precision for some time bins, but calibration provided by the reference stars preserved the relative accuracy of the photometric measurement. 

Telescope jitter caused the targets to move on arcsecond scales among each time binned image. We accounted for this by running \texttt{uvotcentroid} on every time bin image. We then ran \texttt{uvotsource} to perform photometry on the target star with a $3''$ radius circular source and a $27.5 - 35''$ annuluar background region by default.\footnote{Aperture regions were selected in concordance with the suggestions made in the \textsl{Swift}-UVOT Aperture Photometry science thread: https://swift.gsfc.nasa.gov/analysis/threads/uvot\_thread\_aperture.html} 
The background annuli were adjusted for WASP-121, WASP-62, and HAT-P-6 in order to avoid contamination from nearby point sources. All background regions were chosen to cover approximately the same area on the sky as the default background annulus, around $1500$~square-arcseconds.
We also ran \texttt{uvotdetect} on every image, which provided the photometry for our calibration sources.

Next, we defined a set of reference sources to correct for instrumental effects. Using the catalog of aspect correction sources described above, we searched for field sources within $8'$ of our target that were also bright enough to yield high precision photometric results. We filtered for these sources by limiting our consideration to measurements that had $< 3\%$ error and within $7'$ of the target, to avoid the edges of the detector. This process yielded about 3-6 reference sources per target field (Table~\ref{tab:vislist}). We modified this list by hand to account for alternate telescope pointing positions and roll angles, so that every reference source in the final list could be found in every observation.  We identified reference sources in each \texttt{uvotdetect} output file using coordinate matching. The final reference lightcurves were created by computing the error-bar weighted average flux for all reference sources in each time bin. 
We calibrated the target lightcurves by dividing them by the respective reference lightcurves, producing a relative photometric measurement that is corrected for instrument systematics. The target, reference, and final corrected light curves are discussed in more detail in the Appendix.

\subsection{Lightcurve Fitting}
\label{sec:fitting}

We normalized each visit lightcurve for intrinsic stellar variability to first order. 
For each visit lightcurve, we fit a one-dimensional line to the out-of-transit data, using the transit ingress and egress times derived from the optical datasets, and excluding data points that deviated $> 3 \sigma$ from the out-of-transit mean.
We then normalized each lightcurve by the linear model and phase folded all visits using the orbital information from the references listed in Table~\ref{tab:fitlist}, which were supplied to the NEA ephemeris calculator. 

We used the transit simulator, \texttt{BATMAN} \citep{Kreidberg2015}, to fit the normalized, phase folded lightcurves. The free parameters in the transit model are the planet radius relative to the stellar radius ($R_p/R_*$) and the transit midpoint ($T_0$). The latter free parameter was included to account for potential systematic errors in the absolute timing of UVOT, ephemerides uncertainty, or unknown transit timing variations. The remaining \texttt{BATMAN} parameters – the orbital period, semi-major axis, inclination, eccentricity, and longitude of periastron – were held fixed to the literature values (references in Table~\ref{tab:fitlist}). We used the quadratic limb-darkening law
and fixed the limb-darkening coefficients to the values obtained from the Exoplanet Characterization Tool Kit (ExoCTK) Limb Darkening Calculator using the Phoenix ACES model for the GALEX UV bandpass. 
We also modified the BATMAN code to allow for $R_p/R_* < 0$, which is non-physical, but signals the code to model an ``inverse transit'', yielding a light curve that increases in flux instead of decreasing during the transit. Allowing for this possibility reduces the likelihood of spuriously detecting a transit from noise.

\begin{figure}[htp]
    \centering
    \includegraphics[width=1\textwidth]{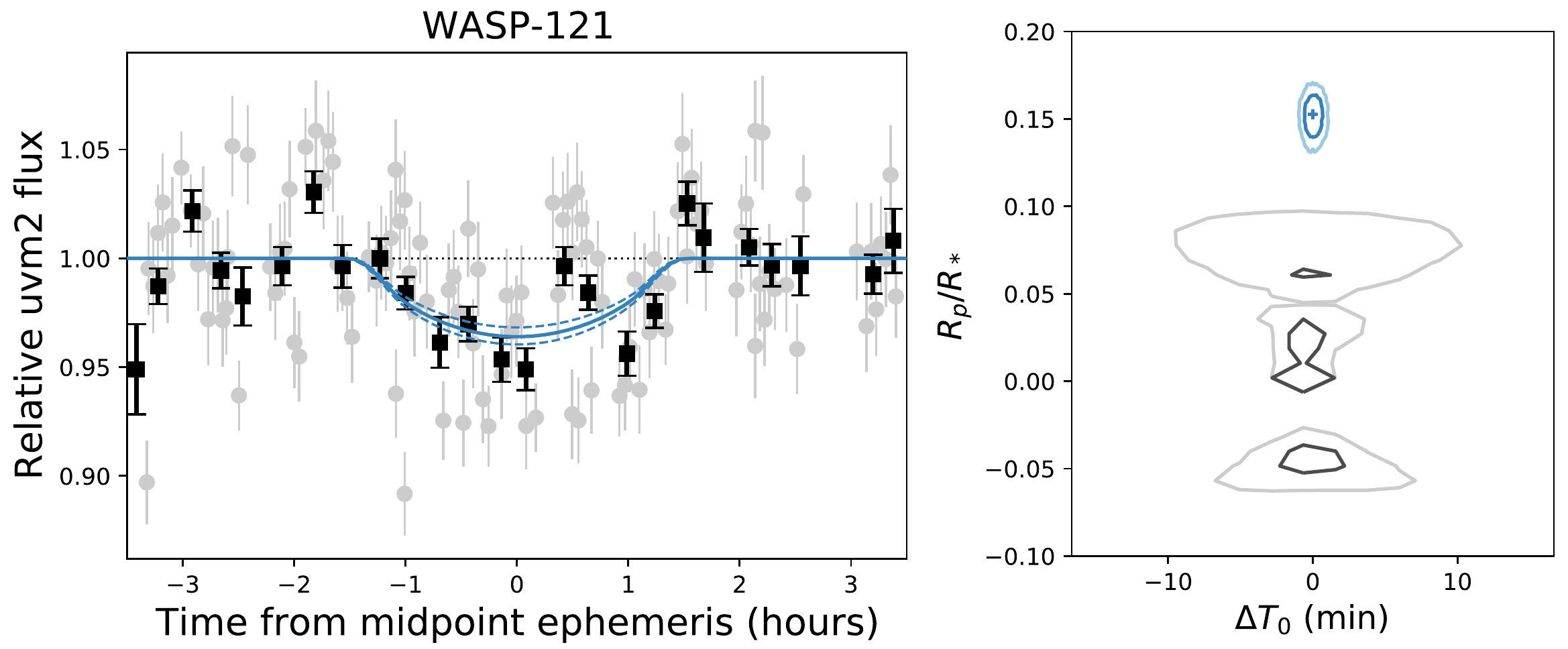}
    \caption{
    (Left) The normalized, phase folded lightcurve for WASP-121 observed with the \textsl{Swift}-UVOT {\sl uvm2} filter. The grey data points show the raw light curve used in fitting. For visual purposes, a light curve binned to 1~ks (16.7~minutes) is over-plotted (black squares). The blue solid curve shows the best-fit \texttt{BATMAN} transit. The blue dashed curves show the $\pm 1\sigma$ confidence intervals on the fit. 
    The transit depth of $2.25 \pm 0.30\%$ is consistent to within 1$\sigma$ of the result obtained by \citet{Salz2019}, who utilized the same dataset. 
    (Right) The blue curves show the 1- and 2-$\sigma$ contours for the posterior distribution derived from the MCMC fit of the WASP-121b transit light curve. The blue cross shows the best fit value. The grey curves show the 1- and 2-$\sigma$ contours for the probability distribution of best-fit values obtained from fitting 100,000 simulated light curves with no transit. 
    $R_p/R_*$ is the planet radius relative to the host star, and $\Delta T_0$ is the offset between the predicted transit midpoint and the fit transit midpoint. In the case of $R_p/R_*<0$, the model light curve is reflected so that an ``inverse transit'' is modelled. The separation between the blue and grey contours was used to assess the statistical significance of the transit detections (Section~\ref{sec:results}).
    }
    \label{fig:W121}
\end{figure}

\begin{figure}[htp]
    \centering
    \includegraphics[width=1\textwidth]{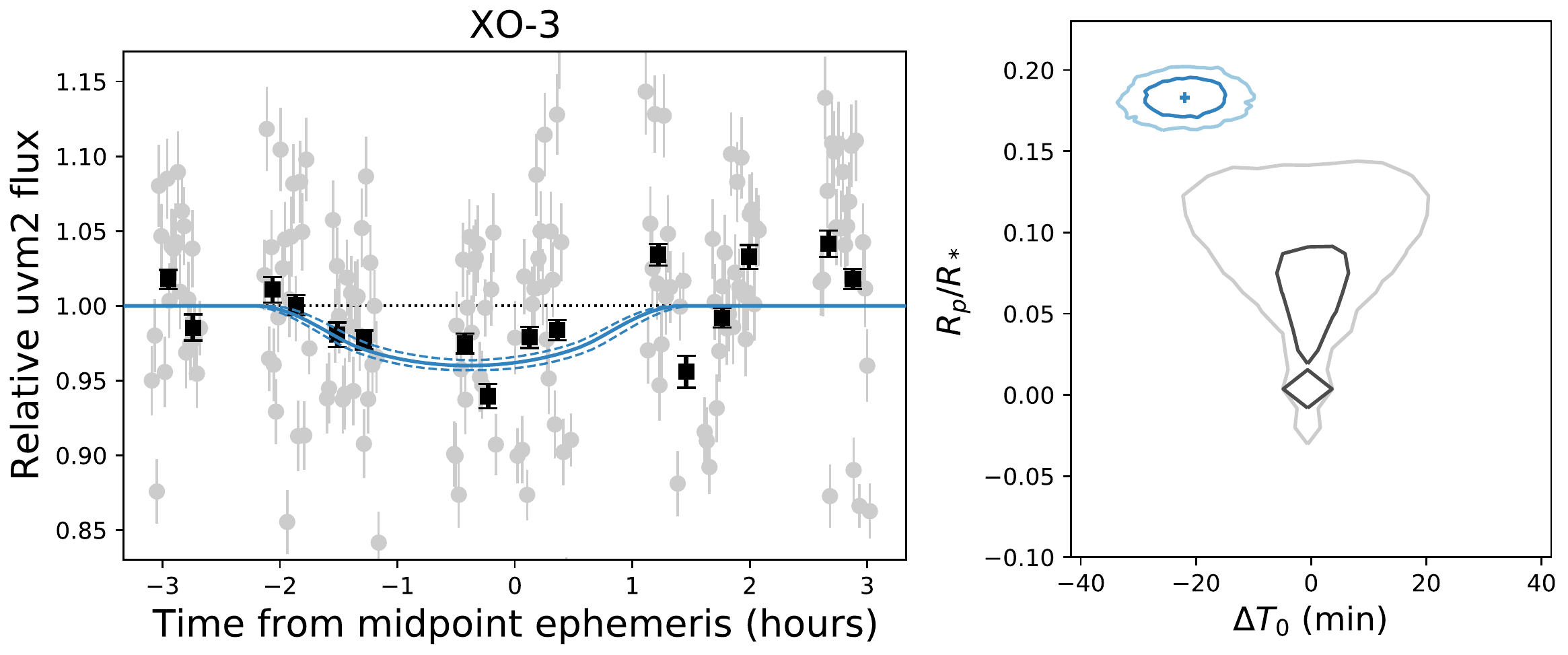}
    \includegraphics[width=1\textwidth]{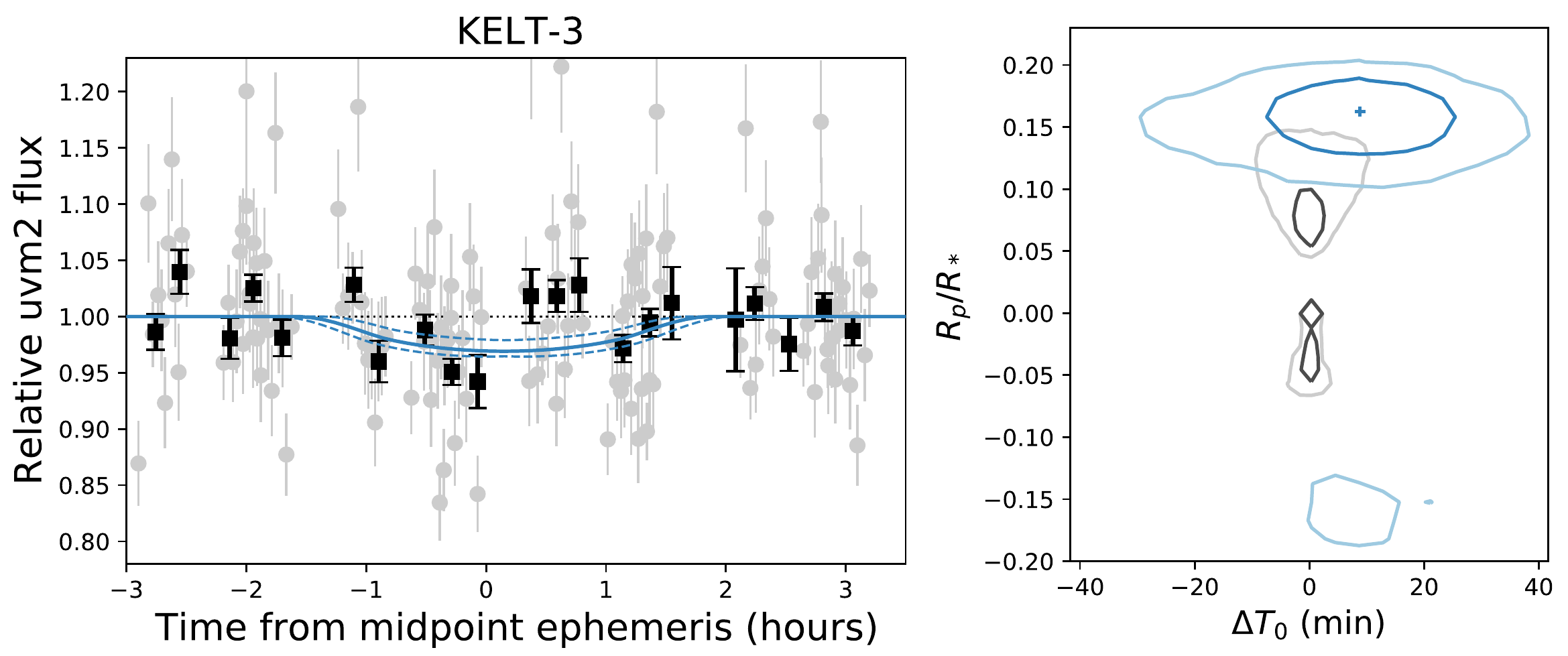}
    \includegraphics[width=1\textwidth]{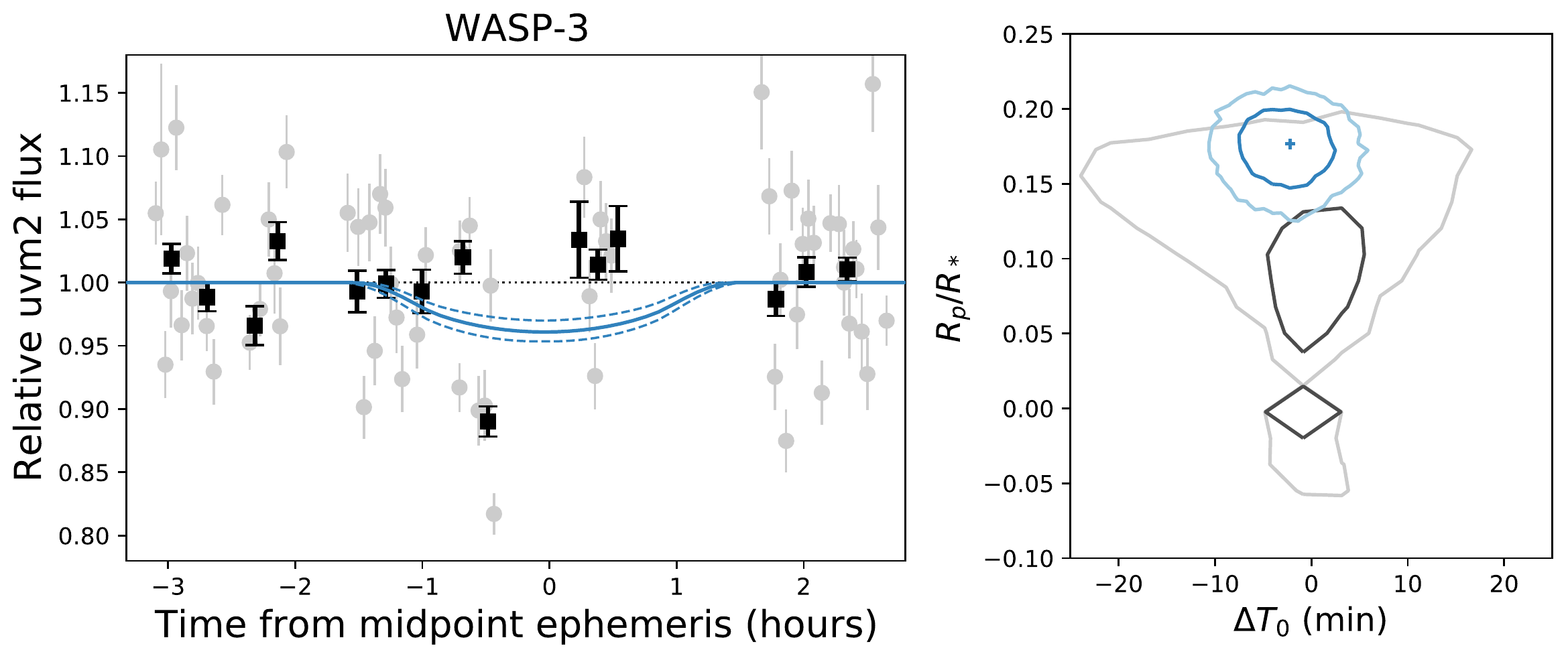}
    \caption{ 
    (Left) The normalized, phase folded lightcurves of the three sources with p-values $\leq 5\%$, using the same plotting conventions as described in Figure~\ref{fig:W121}. 
    (Right) Same contours as in Figure~\ref{fig:W121}. Each transit detection is inconsistent with the null (no transit) hypothesis, with $\geq 1\sigma$ confidence. However, it appears that three datapoints in the light curve of WASP-3b are greatly biasing the transit fit. For the purposes of science interpretation, we adopt the $1\sigma$ null confidence value as the upper limit for the WASP-3b transit, instead.
    }
    \label{fig:good}
\end{figure}

\begin{figure}[htp]
    \centering
    \includegraphics[width=1\textwidth]{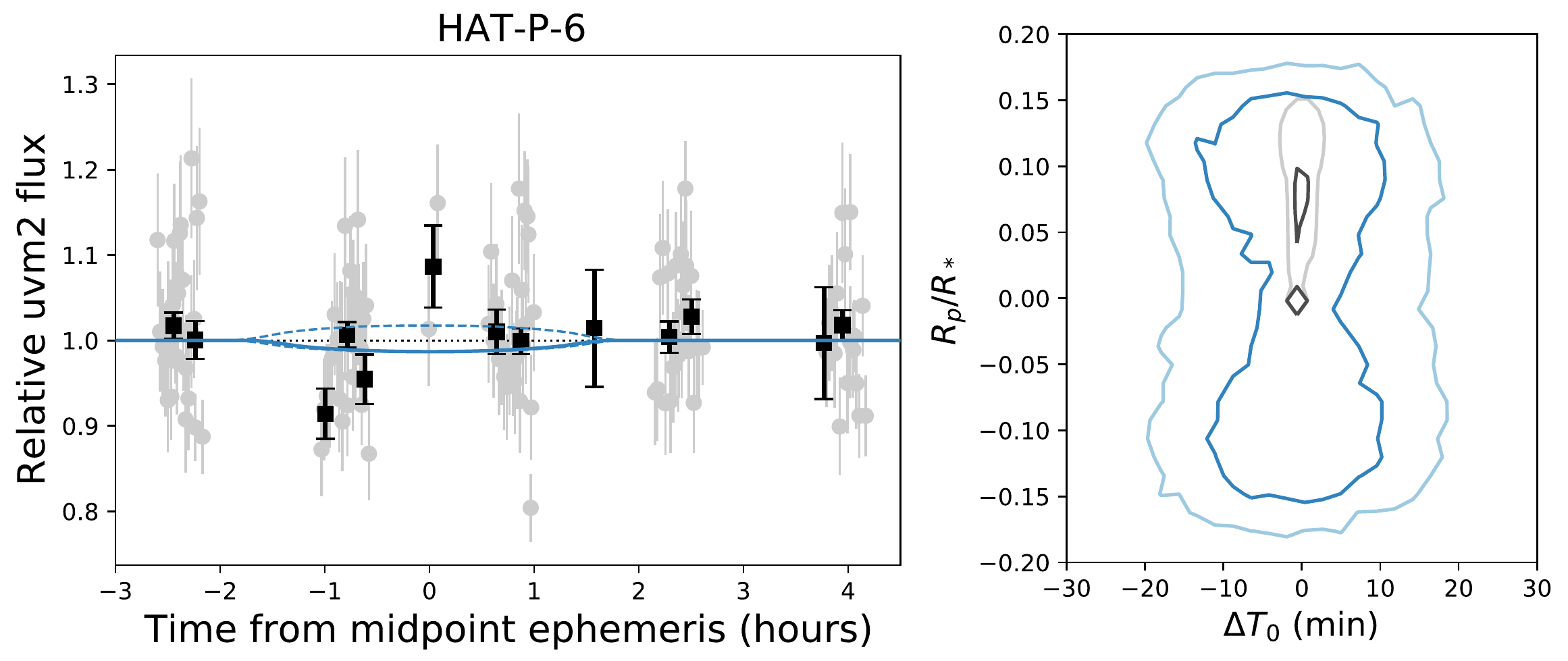}
    \includegraphics[width=1\textwidth]{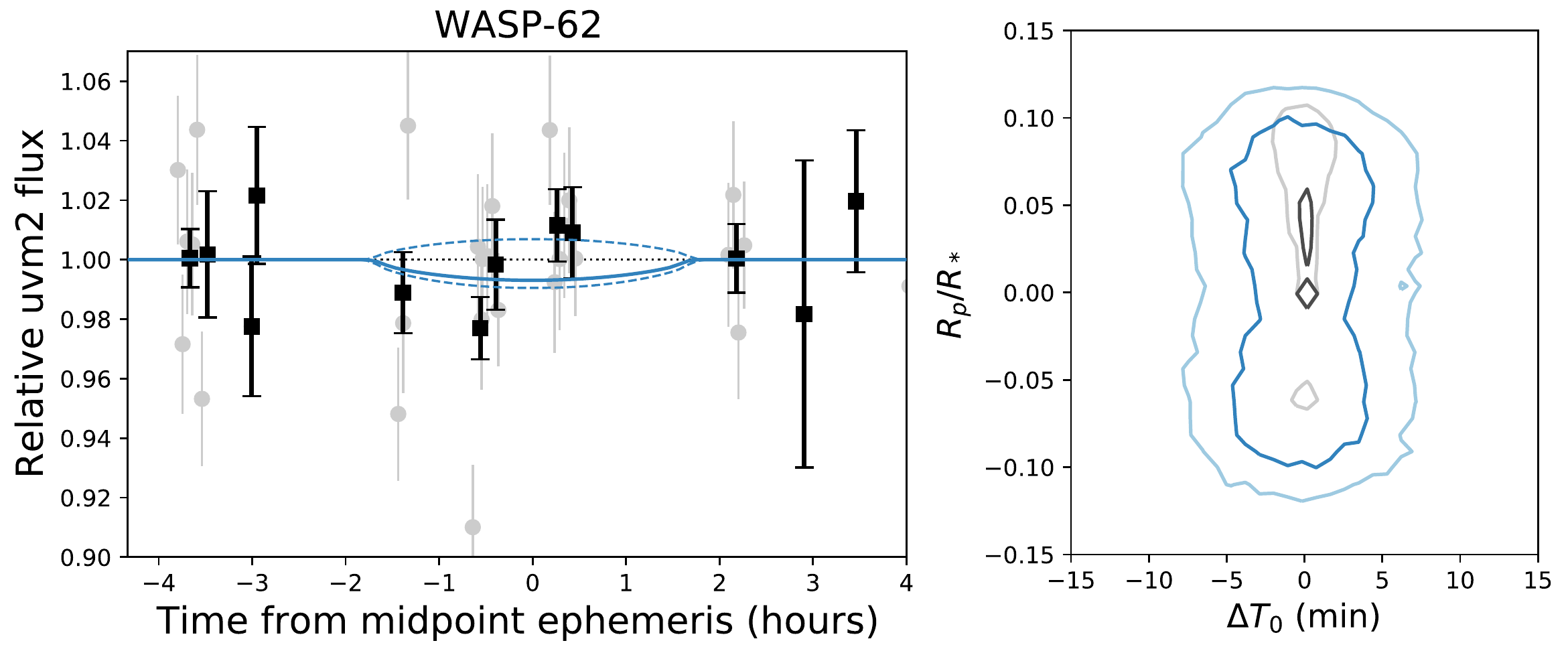}
    \caption{
    (Left) The normalized, phase folded lightcurves of the two sources with no transit detection, using the same plotting conventions as described in Figure~\ref{fig:W121}. 
    (Right) Same contours as in Figure~\ref{fig:W121}.
    }
    \label{fig:notsogood}
\end{figure}

We used the MCMC analysis package \texttt{emcee} \citep{emcee} for fitting. We applied a Gaussian prior to the transit midpoint that was equal to the uncertainty on the ephemeris prediction, obtained from the NASA Exoplanet Archive ephemeris calculator (third column of Table~\ref{tab:fitlist}).
No constraint was applied to the value of $R_p/R_*$. The $\chi^2$ statistic was used to characterize the model likelihood. We ran \texttt{emcee} with 100 walkers, conducted 250 steps for the burn-in phase, then ran the MCMC sampler for 1000 steps to arrive at a final a posterior probability distribution containing $10^5$ sample points. 
Table~\ref{tab:fitlist} (fourth and fifth columns) list the median parameter values with $1\sigma$ confidence intervals, derived from the posterior probability distribution. 
Figures~\ref{fig:W121}-\ref{fig:notsogood} show each transit light curve (grey circles) with the best fit models (blue curves). For visual purposes, a coarser 1~ks (16.7~minute) binned light curve is over plotted (black squares). The binned flux values were computed using the error-bar weighted average of all the data points within that bin.

To quantify the probability that our measured transits arise from natural statistical variation, we performed null hypothesis testing on all targets. % with a measured planet radius $>2\sigma$ from 0. 
All light curves were fit with a ``null hypothesis'' straight-line model with the line position as a free parameter. We then calculated the ratio of the best-fit $\chi^2$ value from the transit model to the best-fit $\chi^2$ value from the null hypothesis (sixth column of Table~\ref{tab:fitlist}). Values $<1$ indicate that the transit model fits better. In order to characterize whether a better fitting transit is due to chance, we derive the $\chi^2$ ratio for a large number of simulated light curves. 
For each target, we simulated a light curve under the null hypothesis by drawing from a normal distribution with the same standard deviation of the real light curve. Error bars were drawn randomly from the set of real light curve error bars. Then we fit the simulated null datasets with a transit model and a null hypothesis model. This process was repeated 100,000 times. 
We then calculated the probability of measuring the observed $\chi^2$ ratio from a light curve that did not contain a transit. This is the p-value (seventh column of Table~\ref{tab:fitlist}), which can be thought of as a false-positive probability for the best fit transit.

\begin{table}[htp]
    \centering
    \caption{Transit model fit results for all targets in this study.}
    \label{tab:fitlist}
    \begin{tabular}{l c c c c c c c} 
        \hline
        \textbf{Target} & \textbf{Opt} & 
        \textbf{$\sigma(\Delta T_0)$} & 
        \textbf{$\Delta T_0$} & \textbf{{\sl uvm2}} & 
        \textbf{$\chi^2$ ratio} &
        \textbf{p-value} & \textbf{NC lim$^\dagger$} \\ 
        & $(\nicefrac{R_P}{R_*})$ &  
        \textbf{Prior} (s) &
        (hours) & $(\nicefrac{R_P}{R_*})$ & (transit/null) & & $(\nicefrac{R_P}{R_*})$ \\
        \hline \hline
        \textbf{WASP-121 b} & $0.12454_{-0.00048}^{+0.00047}$ & 
        25 & $0 \pm 0.006$ &
        $0.15 \pm 0.01$ & 0.856 & $< 10^{-5}$ & 0.10 \\
        \hline
        \textbf{XO-3 b} & $0.09057 \pm 0.00057$ & 
        900 & $-0.37 \pm 0.08$ &
        $0.18 \pm 0.01$ & 0.940 & $0.0007$ & 0.15 \\
        \textbf{KELT-3 b} & $0.0939 \pm 0.0011$ & 
        1700 & $0.14_{-0.18}^{+0.16}$ &
        $0.16 \pm 0.02$ & 0.967 & 0.006 & 0.16 \\ 
        \textbf{WASP-3 b} & $0.1051 \pm 0.0124$ & 
        200 & $-0.04 \pm 0.06$ &
        $0.17 \pm 0.02$ & 0.953 & 0.05 & 0.20 \\
        \textbf{HAT-P-6 b} & $0.09338 \pm 0.00053$ & 
        500 & $-0.01 \pm 0.14$ &
        $0.0 \pm 0.11$ & 0.996 & -- & 0.16 \\
        \textbf{WASP-62 b} & $0.1109 \pm 0.0009$ & 
        200 & $-0.003 \pm 0.06$ &
        $0.0 \pm 0.07$ & 0.997 & -- & 0.11 \\
        \hline 
        \multicolumn{7}{l}{$^\dagger$Null confidence limit, the maximum $R_P/R_*$ value that is $1\sigma$ distinguishable from the noise in each data set.} \\
        \hline
        \multicolumn{7}{l}{References for optical transit depth and transit midpoint ephemeris:} \\
        \multicolumn{7}{l}{XO-3b -- \citet{Winn2008}; KELT-3b -- \citet{Pepper2013}; WASP-3b -- \citet{Christiansen2011} } \\
        \multicolumn{7}{l}{HAT-P-6b -- \citet{Noyes2008}; WASP-62b -- \citet{Hellier2012}; WASP-121b -- \citet{Delrez2016} } \\
        \hline
    \end{tabular}
\end{table}

Finally, we used simulated datasets to compute the $1\sigma$ null confidence limit on the transit depth for all the target lightcurves. 
This process provides an estimate for the largest transit that can hide in a noisy lightcurve.
We simulated 1000 transit lightcurves, using the same standard deviation and error bar distribution as the true lightcurve, for a grid of $R_{p}/R_*$ values ranging from 0 to 0.3. 
Each simulated transit lightcurve was re-fit with a transit model, producing a distribution of best-fit transit radii. The null confidence limit was found by selecting the input $R_{p}/R_*$ value that produced a fit distribution that deviated by $1\sigma$ from the fit distribution derived from the null ($R_{p}/R_* = 0$) simulations. 
The results are listed in the last column of Table~\ref{tab:fitlist}. 
A null confidence limit that is lower than the best fit transit depth  is expected for a positive transit detection. In the case of no transit detection, or if the best-fit transit yielded an $R_p/R_*$ that is less than the null confidence limit (as is the case for WASP-3b), we adopted the null confidence value as the upper limit for $R_p/R_*$.

We applied the analysis techniques described above to the \textsl{Swift}-UVOT dataset for WASP-121b, which has an NUV transit reported in the literature (S19). 
The left side of Figure~\ref{fig:W121} shows the \textsl{uvm2} lightcurve with the best-fit transit model, and the right side shows the posterior distribution from \texttt{emcee} (blue contours) plotted against the best-fit transit models from the simulated null-transit datasets (grey contours). 
The contours demonstrate that, despite the large scatter and poor visual quality of the \textsl{Swift}-UVOT lightcurves, the instrument is capable of measuring $\sim 1\%$ transits at a $> 3\sigma$ level. 
Our transit depth measurement for WASP-121b, $2.25 \pm 0.30\%$, is within 1-sigma of the transit depth reported by S19. 

\subsection{X-ray and UV characteristics of the host stars}
\label{sec:XUV}

We used all available \textsl{Swift} data to characterize the X-ray and UV properties of the exoplanets' host stars. 
The X-ray telescope (XRT) on \textsl{Swift} ran concurrently in photon counting mode throughout all of our UVOT observations. 
We stacked all available XRT data, including observations that did not cover exoplanet transits, summarized in Table~\ref{tab:XUV}. 
We used \texttt{XSELECT} to extract an image and a spectrum from the stacked event file, after filtering for the 0.2-2.4 keV range. 
We then ran the source detection algorithm \texttt{XIMAGE} to search for point sources coincident with the coordinates of our target exoplanet host stars.

The only host star detected in the \textsl{Swift}-XRT datasets was WASP-62, with a $2.5 \sigma$ significance, registering $5 \times 10^{-3}$ counts per second. Because there was not a sufficient number of counts to fit a spectrum, we used the CXC webtool, \href{https://cxc.harvard.edu/toolkit/pimms.jsp}{PIMMS v4.10}, to convert the \textsl{Swift}-XRT count rates to a flux value. 
Using the Gaia g-band extinction value of 0.1843 \citep{GaiaMission, GaiaDR2} and the scaling relations for ISM column versus visual extinction \citep{Foight2016}, we arrive at a value of ${\rm N}_{\rm H} = 5.3 \times 10^{20}$~cm$^{-2}$ for WASP-62. 
We supplied this ${\rm N}_{\rm H}$ value to model the extinction of a 0.3~keV temperature APEC model in PIMMS, yielding an absorbed flux of $(1.0 \pm 0.5) \times 10^{-14}$~erg~cm$^{-2}$~s$^{-1}$ and an unabsorbed flux of $(1.7 \pm 0.8) \times 10^{-14}$~erg~cm$^{-2}$~s$^{-1}$. 

We also examined the publicly available XMM-Newton data for WASP-62
(PI:~Sanz-Forcada). Obtaining a 46~count spectrum for the source, we fit a single temperature APEC model \citep{APECmodel} together with a TBABS model for interstellar absorption \citep{Wilms2000}. We binned the spectrum to a minimum of five counts per bin and used the C-statistic \citep{Cash1979} as our fit statistic. The errors on the fitted values are evaluated using Xspec's built-in MCMC sampler. 
We obtain a 0.2-2.4~keV absorbed (unabsorbed) flux value of $1.38_{-0.20}^{+0.28}\ (2.51_{-0.36}^{+0.51}) \times 10^{-14}$~erg~cm$^{-2}$~s$^{-1}$. 
With the absorption correction, we find an intrinsic luminosity that is 50\% larger than that reported in \citet{Alam2020}. 
Because there is significantly more X-ray information in the XMM-Newton dataset, we use this value in Table~\ref{tab:XUV} and in subsequent analysis.

For the remaining exoplanet host stars, no point sources were detected. We used the background count rates, scaled to a $10''$~radius region, to estimate the upper limit to the absorbed flux for each point source.
We again used PIMMS to convert the count rates to flux, modeling the stellar coronae with a 0.3~keV APEC model with solar abundances. As with WASP-62, we used the Gaia g band extinction and ISM scaling relations to estimate ${\rm N}_{\rm H}$. The ISM columns were used to compute an upper limit to the unabsorbed flux, and subsequently luminosity, for each exoplanet host star. The results are summarized in the last three columns of Table~\ref{tab:XUV}.

Finally, 
we ran \texttt{uvotsource} on every \textsl{uvm2} image produced by the standard \textsl{Swift} pipeline, and computed the error weighted average. The results are listed in the last column of Table~\ref{tab:XUV}. Nearly all of the measured magnitudes are similar to those available in the GALEX catalog (listed in Table~\ref{tab:vislist}), except for XO-3, which appears a factor of three brighter at the time of the \textsl{Swift} measurements.

\begin{table}[htp]
    \centering
    \caption{XUV properties of the host stars}
    \label{tab:XUV}
    \begin{tabular}{l c c c c c} 
        \hline
        \textbf{Target} 
        & \textbf{X-ray Exp}
        & \textbf{$\log F_X({\rm abs}) ^\dagger$} 
        & \textbf{${\rm N}_{\rm H}\ ^\ddag$}
        & \textbf{$\log L_X ^\dagger$} & \textbf{UV} \\
        & [ks] & [erg/s/cm$^2$] & [$10^{21}$~cm$^-2$] & [erg/s] & AB mag \\
        \hline \hline
        \textbf{XO-3} & 13.8 & $< -14.6$ & 1.7 & $< 28.6$ & $13.5 \pm 0.02$ \\ 
        \textbf{KELT-3} & 33.9 & $< -14.9$ & 2.2 & $< 28.4$ & $14.2 \pm 0.02$ \\
        \textbf{WASP-3} & 12.4 & $< -14.1$ & 0.58 & $< 29.0$ & $14.7 \pm 0.02$ \\
        \textbf{HAT-P-6} & 16.0 & $< -14.6$ & 2.3 & $< 28.9$ & $14.0 \pm 0.02$ \\
        \textbf{WASP-62} & 5.2 (XMM) & $-13.86^{+0.09}_{-0.06}$ & 0.53 & $28.96^{+0.09}_{-0.06}$ & $15.0 \pm 0.02^*$ \\
        \hline 
        \hline
        \multicolumn{6}{l}{$^\dagger$0.2-2.4~keV band} \\
        \multicolumn{6}{l}{$^\ddag$Using ISM scaling relations of \citet{Foight2016}} \\
        \multicolumn{6}{l}{$^*$NUV magnitude derived from total $12.5$~ks \textsl{Swift}-\textsl{uvm2} dataset} \\
        \hline
    \end{tabular}
\end{table}

\section{Results} \label{sec:results}

In this section, we review the transit fitting results for each target and describe what the NUV measurements add to our understanding of each system. We start with the most likely NUV transit candidates and end with the null detection systems. For each target, we examine the planetary radius relative to its Roche lobe in order to determine the likelihood that the NUV transit is probing an escaping atmosphere. 
The size of the Roche lobe is approximated by following \citet{Eggleton1983}:
\begin{equation}
\label{eq:roche}
    \frac{R_{\rm RL}}{a} = \frac{0.49q^{2/3}}{0.6q^{2/3}+ \text{ln}(1 + q^{1/3})}
\end{equation}
where $r$ is the radius of a sphere that approximates the Roche lobe for a planet, $a$ is the semi-major axis of the orbit, and $q = \nicefrac{M_p}{M_*}$ where $M_p$ is the planetary mass and $M_*$ is the stellar mass. 
For all of our targets, we use the radial-velocity measurements for the masses of both the planet and the host star provided by \citet{Bonomo2017}. 
When converting $R_P/R_*$ to Jupiter radii, we use $R_\odot = 6.957 \times 10^{8}$~m for the radius of the Sun and $R_J = 7.1492 \times 10^{7}$~m for the radius of Jupiter.

We also apply the scaling relations of \citet{King2018} to estimate the total XUV
(13.6~eV - 2.4~keV) 
luminosity of the host star from the 0.2-2.4~keV band. This allows us to estimate the mass-loss rate from photoevaporation in the energy-limited regime:
\begin{equation}
    \dot{M} = \beta^2 \eta \frac{\pi F_{XUV}{\rm (planet)} R_p^3}{G K M_p}
\end{equation}
where
\begin{equation}
    K = 1 + \frac{3}{2 \xi} + \frac{1}{2 \xi^3}
\end{equation}
and $F_{XUV}$ is the XUV flux incident on the planet, $\beta \equiv R_{XUV} / R_p$ is the ratio between the planet radius at XUV wavelengths compared to the planet surface radius, $\xi \equiv R_{\rm RL} / R_p$, and $\eta$ is the efficiency factor for converting irradiation to mass-loss, which we assume is $0.1$ \citep{Watson1981, Erkaev2007}. 
We use the stellar radius provided by the Gaia~DR2 catalog \citep{GaiaMission, GaiaDR2} in order to compute $R_p$(opt) and $\xi$. We assume $\beta = 1$ as a benchmark value. We expect the true $\beta$ value for our systems to range $\sim 1-2$, which would affect the mass-loss estimates by a factor of four. 
This is comparable to our uncertainty in the true value for $\eta$. 
Table~\ref{tab:comparison} summarizes the conclusions from this section.

\begin{table}[htp]
    \centering
    \caption{Exoplanet properties derived in this work}
    \label{tab:comparison}
    \begin{tabular}{l c c c} 
        \hline
        \textbf{Target} & \textbf{Mass loss rate} &
        $R_*^\dagger$ & $R_P$(NUV) \\ 
        & (g s$^{-1}$) & ($R_\odot$) & $(R_J)^\ddag$ \\
        \hline \hline
        \textbf{XO-3 b} & $< 2 \times 10^{10}$
        & $ 1.452 \pm 0.098 $ & $2.54 \pm 0.22$ \\
        \textbf{KELT-3 b} & $< 1.5 \times 10^{11}$
        & $ 1.604_{-0.136}^{+0.058} $ & $2.50_{-0.38}^{+0.33}$ \\
        \textbf{WASP-3 b} & $< 2 \times 10^{11}$
        & $ 1.299_{-0.075}^{+0.063} $ & $ \leq 2.53 $ \\
        \textbf{HAT-P-6 b} & $< 3 \times 10^{11}$
        & $ 1.610_{-0.157}^{+0.041} $ & $ \leq 2.51 $ \\
        \textbf{WASP-62 b} & $5 \times 10^{11}$ 
        & $ 1.299_{-0.153}^{+0.064} $ & $ \leq 1.39 $ \\
        \hline
        \multicolumn{4}{l}{$^\dagger$Gaia Data Release 2 \citep{GaiaDR2}} \\
        \multicolumn{4}{l}{$^\ddag$Adopting the values of $R_{\odot} = 6.957 \times 10^{10}$~cm and $R_J = 7.149 \times 10^9$~cm} \\
        \hline
    \end{tabular}
\end{table}

\subsection{XO-3}

XO-3, a main-sequence F5V star, has one of the most unique planetary systems ever measured. XO-3b is particularly massive ($11.7 \pm 0.4$~M$_{\rm J}$),  has a large eccentricity ($e = 0.28$), and 
its orbital plane is highly inclined ($\lambda = 37.3^\circ$) relative to the equatorial plane of its host star \citep{Johns-Krull2008, Winn2008, Hebrard2008, Winn2009, Bonomo2017}.
The high eccentricity and orbital inclination, thought to arise from planet-planet interactions early in the system's history, makes XO-3b an important subject for understanding the role of tidal forces in planetary evolution.

\textsl{Swift} conducted two visits of the XO-3b transits, three months apart. % (obsIDs 001-009 \& 010-018). %
Despite large systematic variance in the lightcurve amplitude, the average appears to decrease around the expected transit time. Null hypothesis testing indicates that this is the most likely transit in our sample, with a very low likelihood of this being a false detection. Based on the p-value and contours in Figure~\ref{fig:good}, we consider this a $3\sigma$ detection.

It is notable that the modeled XO-3b transit midpoint is more than $3\sigma$ away from the midpoint predicted from the \citet{Winn2008} ephemeris. 
However, the 22~minute offset is close to the propagated ephemeris uncertainty of 0.01~days (14~min). 
We experimented with a few different ephemeris values from the literature \citep{Wong2014, Bonomo2017} and found that all of them returned an early NUV transit for XO-3b ($\Delta T_0 = -0.47$~hours and $-0.49$~hours, respectively). Most importantly, none of them significantly altered the transit depth of XO-3b; both alternative ephemerides returned $R_p/R_* = 0.20 \pm 0.01$. Future simultaneous optical observations of an XO-3b transit can be used to determine whether or not the NUV transit is centered with respect to the optical transit.

Past studies on the XO-3b system suffered from large uncertainties due to the unknown distance to XO-3, which can now be resolved. Using the Gaia~DR2 values for the stellar radius, the transit depth measured by \citet{Winn2008} implies that XO-3b is $1.280 \pm 0.087~R_{\rm J}$. Using the stellar distance provided by Gaia~DR2, the optical flux of the star on the planet surface (insolation) is $2.8 \times 10^{9}$~erg~s$^{-1}$. 
We use these values with the most up-to-date mass and orbital parameters inferred from radial velocity measurements \citep{Bonomo2017} to estimate the ratio of tidal heating versus energy available to photoabsorptive heating, $\dot{E}_{\rm tide}/\dot{E}_{\rm insolation}$, described in \citet{Machalek2010} and references therein. We find $\dot{E}_{\rm tide}/\dot{E}_{\rm insolation} = 0.44 (Q_p/10^5)^{-1}$, where $Q_p$ is the dimensionless heating parameter. To explain a radius of $1.28~R_{\rm J}$, XO-3b would require a $Q_p \approx 10^6$ \citep{Liu2008}, implying a non-negligible amount of tidal heating: $\sim 5\%$ of luminosity incident on the planet surface.
This formula does not take into account the efficiency of photoabsorption in the planet's atmosphere. Assuming a nominal value of 10\% heat transfer efficiency, one can conclude that the atmospheric heating due to tidal forces on XO-3b is roughly equal to the heating from stellar irradiation.

With a transit depth of $3.2 \pm 0.4\%$, our measurement of XO-3b also has the largest value for NUV radius relative to the host star, in our sample. 
The particularly high mass of XO-3b should prevent it from experiencing significant mass-loss. The NUV radius, $R_p$(\textsl{uvm2})$ \; = 2.54 \pm 0.22~R_{\rm J}$, is well within the planet's Roche lobe of $9.4~R_{\rm J}$.
Given the upper limit of the X-ray flux, we find that the total EUV luminosity of the star is $< 7.5 \times 10^{29}$~erg~s$^{-1}$ and the predicted mass-loss for the system is $< 2 \times 10^{10}$~g~s$^{-1}$. 
This is the lowest mass-loss rate calculated for the planets in our sample.

\subsection{KELT-3}

KELT-3b is a $1.5~M_{J}$ hot Jupiter with an optical radius of 1.3~$R_J$ (or $1.5~R_J$ using the stellar radius from Gaia DR2), orbiting an F7V-type star \citep{Pepper2013}. 
The insolation received by the planet, $10^9$ $\text{erg s}^{-1}$ $\text{cm}^{-2}$, 
places it in the class of hot Jupiters observed have ``inflated'' radii \citep{Demory2011}.

We obtained two visits of the KELT-3b transit in 2019. Preliminary analysis of that work hinted at a transit, but an anomalous flare appeared to contaminate the light curve \citep{Ravi2020}. However, KELT-3 does not exhibit strong CaII H and K lines, which are positive indicators of stellar activity \citep{Pepper2013}. In this work, we found that normalizing the KELT-3 light curve by the averaged light curve of our reference sources removed the apparent flare, demonstrating that our techniques are sufficient to remove instrumental variations from the light curve (see Appendix). 
To compliment this conclusion, we obtained a subsequent transit observation in 2020 (Visit 3) to fill sample gaps in the original light curve.

Our best-fit transit depth is 2.6 $\pm$ 0.6\%, which is 70\% deeper than the optical transit. Our p-value of 0.7\% implies a 2.5$\sigma$ detection, but visual inspection of the fit contours (Figure~\ref{fig:good} right) suggest that it is only distinguishable from the null hypothesis by about 1.5$\sigma$.

The Roche lobe size for KELT-3b is 4.3~$R_J$, which is nearly a factor of two larger than its NUV radius of $2.5 \pm 0.3~R_J$. 
A search for H$\alpha$ emission as a signature for an extended or escaping atmosphere from KELT-3b yielded a non-detection \citep{Cauley2017}. They place an upper limit on the size of the H$\alpha$ absorbant atmosphere that is 10\% larger than the optical planet disk. However, the lack of a H$\alpha$ signature is most likely a sign that $< 0.1\%$ of the hydrogen in the KELT-3b atmosphere is in an excited state, which can be expected in planetary systems older than 1.2~Gyr \citep{Allan2019}. 
The EUV flux incident on the planet is $< 5 \times 10^{29}$~erg~s$^{-1}$, and the estimated mass-loss rate is $< 1.5 \times 10^{11}$~g~s$^{-1}$. This places KELT-3b in a moderate mass-loss regime for our sample, even though it does not have a visibly escaping atmosphere.

\subsection{WASP-3}

WASP-3b is a 1.9~$M_J$ planet with 1.4~$R_J$ radius orbiting a type F7-8V star with the very short period of 1.85~days, making it one of the the most irradiated hot Jupiters at the time of its discovery \citep{Pollaco2008, Bonomo2017}. 
Eclipse measurements with \textit{Spitzer} revealed dayside temperatures $2200 - 2400$~K, placing it in the category of ultrahot Jupiters, and best-fit models over the $2-10~\mu{\rm m}$ spectral energy distribution imply that WASP-3b has an inverted temperature profile with ineffcient heat exchange between the day and night sides of the planet \citep{Zhao2012, Rostron2014}. 
The ultra-hot status of WASP-3b makes it similar to WASP-121b, which exhibits a range of dayside temperatures from $2300-2800$~K \citep{Garhart2020, Bourrier2020}. 
By virtue of its similarity to WASP-121b, WASP-3b is the top planet in our sample from which  one might expect the deepest NUV transit.

We searched for an NUV transit from three visits of \textsl{Swift} taken over the course of one month. %
Our best-fit transit depth, $3.2 \pm 0.7\%$, has a p-value of 5\%, implying a marginal $<2 \sigma$ detection. This conclusion is also supported by the contours in Figure~\ref{fig:good} (right). 
However, in examining the light curve in Figure~\ref{fig:good} (left), it appears that the transit fit is biased by three data points, which show a 5-10\% lower flux measurement, all of which arise from a single exposure within transit Visit 3 (see Appendix). We are unable to attribute the anomalously low flux measurement to any obvious artifact in the image file. Based on the lack of obvious transit signatures from the remaining light curve, we adopt the $1\sigma$ null confidence value as the upper limit to the WASP-3b transit depth.

The corresponding upper limit for the radius of WASP-3b is $2.5~R_J$, which is smaller than its Roche lobe radius, 3.6~$R_J$. 
The upper limit to the X-ray flux of WASP-3 is consistent with, and more restrictive than, a previous X-ray brightness limit reported by \citet{Kashyap2008}. We calculate an EUV luminosity $< 8 \times 10^{29}$, and a corresponding evaporative mass-loss rate of $< 2 \times 10^{11}$~g~s$^{-1}$. This is consistent with the evaporation rate of $4 \times 10^{11}$~g~cm$^{-1}$, predicted by \citet{Bourrier2015}, who assumed $\eta = 0.3$ for the heating efficiency.

\subsection{HAT-P-6}

HAT-P-6b has an optical radius of approximately $1.5~R_J$, a mass of $1.1~M_J$, and an optical insolation of $2 \times 10^{9}$~erg~cm$^{-2}$~s$^{-1}$, placing it in the class of inflated hot Jupiters \citep{Noyes2008, Bonomo2017, Demory2011}. 
\textsl{Swift} made two visits of the HAT-P-6b transit, but the planet  was not detected (Figure~\ref{fig:notsogood}, right). The NUV transit upper limit of $R_p/R_* < 0.16$ implies a radius $R_{\rm NUV} < 2.5~R_J$. 
The Roche lobe for this system is approximately $7~R_J$, substantially larger than NUV radius. The observational limit on the X-ray flux implies an EUV luminosity $< 10^{30}$~erg~s$^{-1}$, which yields an estimated mass-loss rate $< 3 \times 10^{11}$~g~s$^{-1}$.

\subsection{WASP-62}

WASP-62b orbits a late F-type star and exhibits the largest optical transit depth of all targets in this work. 
Its large radius of $1.4~R_J$, low mass of 0.6~$M_J$, and optical insolation of $10^{9}$~erg~cm$^{-2}$~s$^{-1}$ puts WASP-62b in a curious category of low-density inflated planets \citep{Hellier2012, Brown2017, Bonomo2017}. 
The young age of this planetary system, $\sim 0.4 - 1.3$~Gyr based on gyrochronology \citep{Hellier2012}, suggests that heating via XUV radiation, which dominates younger systems \citep{King2020}, might also be important for explaining this planet's low density appearance.

\textsl{Swift} visited WASP-62 three times over the course of a month. However, the first visit 
missed the transit of WASP-62b by about four hours. 
Visits 2 and 3, unfortunately, sparsely sample the transit. Relatively few data points are available outside of the transit window, making it particularly difficult to obtain a baseline lightcurve for the host star. 
We attribute these factors to our null result, demonstrated in Figure~\ref{fig:notsogood} (right). 
This dataset provides us with an upper limit of $R_p/R_* < 0.11$ ($R_{\rm NUV} < 1.4~R_J$), which is consistent with the optical radius of WASP~62b.

The Roche lobe radius for WASP-62b is $6~R_J$, a factor of \textbf{4} larger than the upper limit to the NUV radius. 
As described in Section~\ref{sec:XUV}, WASP-62 was the only host star in our sample with a positive X-ray detection. 
The estimated EUV luminosity of the host star is $(1.4 \pm 0.2) \times 10^{30}$~erg~s$^{-1}$, and the estimated mass-loss rate is $5 \times 10^{11}$~g~s$^{-1}$. This is a factor of two larger than the value predicted by \citet{Bourrier2015}.

\section{NUV transit measurements in context} 
\label{sec:RNUVcompare}

In this section, we evaluate the ratio of the NUV and optical planet radius ($R_{\rm NUV}/R_{\rm opt}$) as a way to probe atmospheric properties. 
We calculated this ratio by dividing the square-root of the transit depth ($R_{P}/R_{*}$) observed in the NUV versus that in the optical (Table~\ref{tab:fitlist}). In the $200-300$~nm range, the NUV and optical stellar radius are expected to be identical. Limb brightening effects from optically thin outer layers, which would make the star appear larger, are not apparent until EUV wavelenghts \citep[e.g.,][]{Llama2015}. Thus, our measurement of ($R_{\rm NUV}/R_{\rm opt}$) controls for uncertainty in the host star radius.

We used the radiative transfer code Exo-Transmit \citep{Kempton2017} to assess whether the NUV transit depths could be explained by Rayleigh scattering, which leads to an increase in the apparent radius of the planet at shorter wavelengths. 
We generated transmission spectra for each of our targets, assuming solar metallicity and using the temperature-pressure profiles built into the code, set equal to the equilibrium temperature for each planet. 
Because the Exo-Transmit opacity database cuts off at 300~nm, we  extrapolated the Rayleigh scattering slope to NUV wavelengths. 
We varied the Rayleigh factor, cloud deck pressure, and planet radius to examine how much the NUV planet radius could vary while still fitting the optical data. 
We found no combination of parameters that could produce $R_{\rm NUV}/R_{\rm opt}$ values larger than 1.05, which is too small to explain some of the positive NUV transit detections, which have $R_{\rm NUV}/R_{\rm opt} \sim 1.2-2$. 

\subsection{NUV transits as a probe of atmospheric mass-loss}
\label{sec:MassLoss}

\begin{figure}[tp]
    \centering
    \includegraphics[width=.49\textwidth]{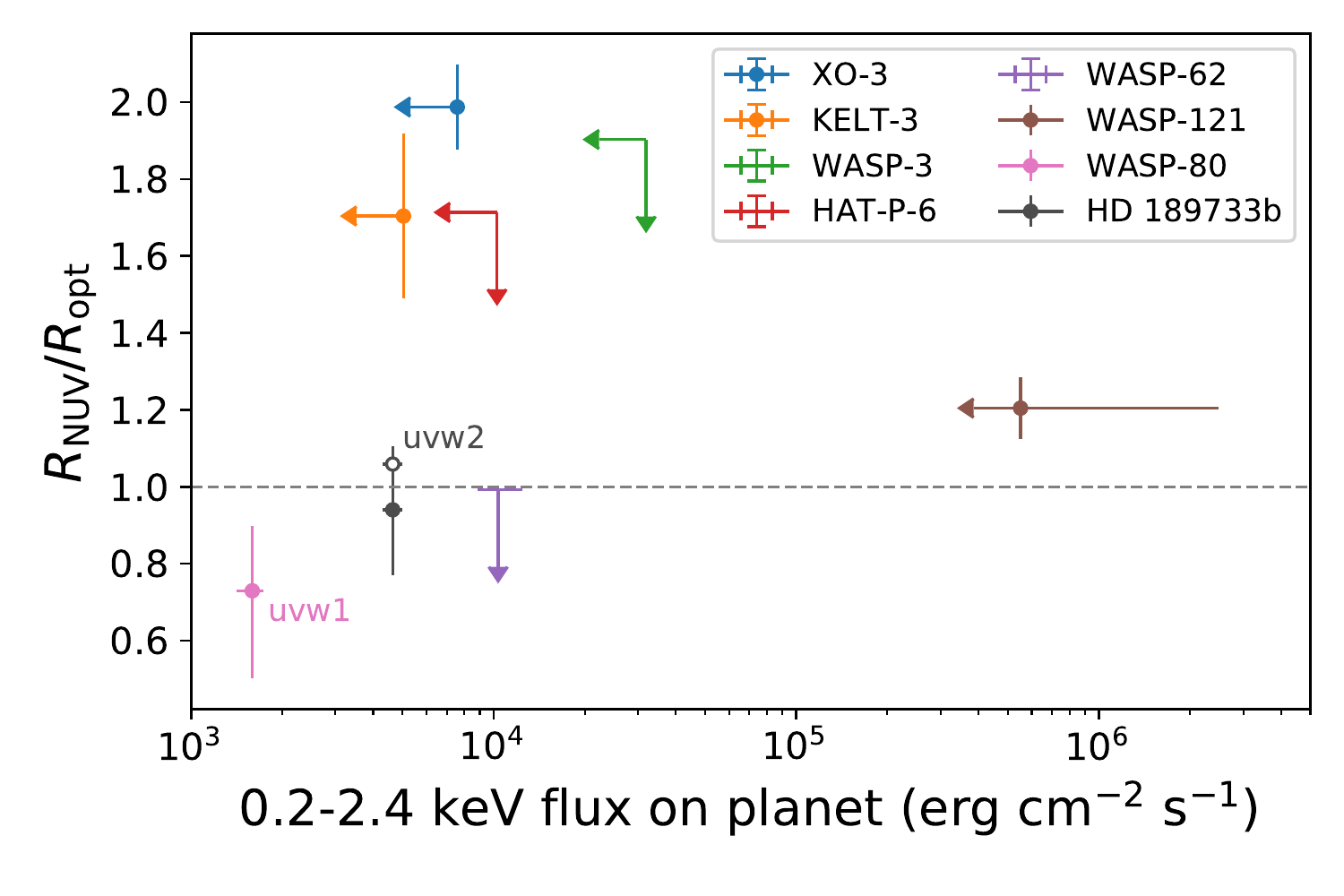}
    \includegraphics[width=.49\textwidth]{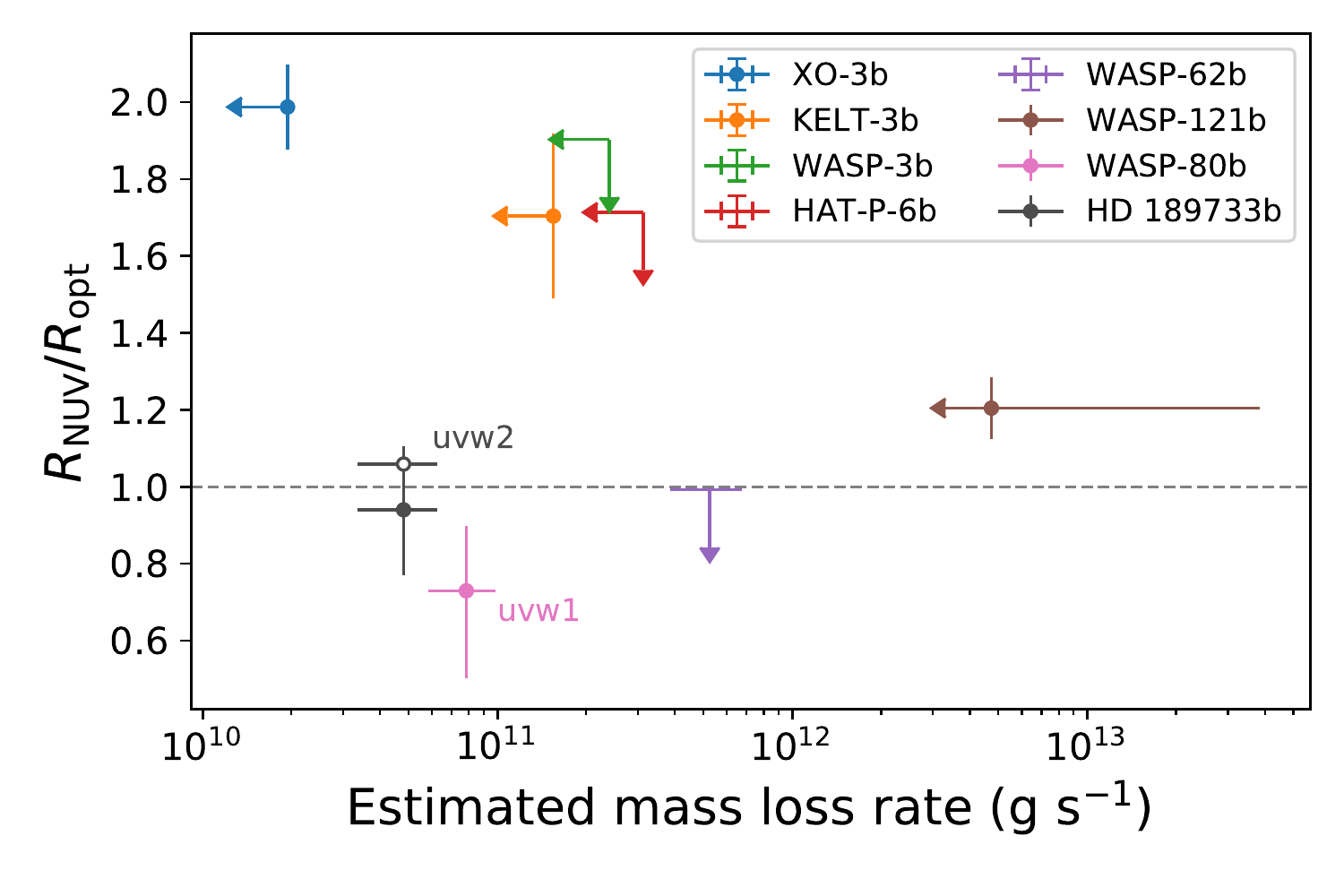}
    \caption{(Left) The ratio of the NUV and optical radius versus the 0.2-2.4~keV X-ray insolation. The arrows signify upper limits arising from either a non-detection in the X-ray flux or from the upper limit derived from the NUV transit depth. All measurements were obtained with a \textit{uvm2} filter except where labeled: the WASP-80b result was obtained with  \textit{uvw1} and HD~189733b (open circle) with \textit{uvw2}.
    (Right) The same ratio plotted against the mass-loss rate for each planet, which was derived from the X-ray luminosities of the host stars.
    }
    \label{fig:insolation}
\end{figure}

Here, we test the hypothesis set forward by S19, that NUV transit depth provides a probe of atmospheric escape. One of the leading theories to explain atmospheric mass-loss is photoevaporation, which is dominated by the extreme UV (EUV, $40-912$ \AA). 
In Section~\ref{sec:analysis}, we utilized the X-ray luminosity of the host star to estimate the EUV component and corresponding mass loss by photoexcitation  processes. However, the X-ray luminosity of a host star might also correlate with other evaporative processes, such as mechanical erosion by stellar winds and coronal mass ejections. In this section, we look for correlations in $R_{\rm NUV}/R_{\rm opt}$ for the observed characteristic (X-ray irradiation) and for the derived characteristic (photoevaporative mass loss rate).
We calculated the X-ray insolation of each planet by scaling the unabsorbed luminosity to the flux at the semi-major axis of the planetary orbit. 
In four out of our five cases, we are only able to derive upper limits. 

In this examination we also include WASP-80b \citep{King2018}, WASP-121b (S19), and HD~189733b \citep{King2021} for which positive NUV transit detections have been derived and X-ray measurements of the host stars exist.
The X-ray insolation for HD~189733b is derived from the X-ray luminosity measurement of \citep{Poppenhaeger2013}. 
We treat the X-ray measurement of WASP-121 as the upper range to the stellar luminosity, because S19 reported that X-ray emission from the star was only detected in a subset of the \textsl{Swift} observations. The measurement likely characterizes the star in a flare state, not its baseline level of activity.
We also plot the WASP-121b NUV measurement derived in this work (Table~\ref{tab:fitlist}) rather than the S19 value. 
Finally, note that the NUV transit depth for WASP-80b was obtained with the XMM-Newton Optical Monitor (OM) instrument, which has the same filters as the \textsl{Swift}-UVOT. The WASP-80b datapoint differs from the other measurements in two significant ways. First, the lightcurve was obtained using a different instrument operation mode (fast window) and analysis techniques. Second, the data were taken with the \textsl{uvw1} filter, which covers a redder wavelength range than \textsl{uvm2}.

Figure~\ref{fig:insolation} (left) shows the ratio of the NUV to optical planet radius as a function of X-ray insolation. 
The estimated photoevaporative mass loss rate is plotted in Figure~\ref{fig:insolation} (right). For host stars with positive X-ray detections, WASP-80 and WASP-62, we injected a 50\% uncertainty on the value of $\eta$ to account for uncertainty in both the XUV flux derived from the 0.2-2.4~keV measurement and the uncertainty in the mass loss efficiency. 
The energy limited mass loss rate for HD~189733b is taken from \citet{Kubyshkina2018}, assuming a 30\% uncertainty.
The radius derived from NUV transit measurements does not appear to be positively correlated with the mass-loss rate.

It is difficult to interpret the large NUV transit depths observed for XO-3b and potentially KELT-3b under the current paradigm of atmospheric escape.
Rudimentary cloud-free models for the UV transmission spectra of exoplanet atmospheres have demonstrated that FeII is the dominant gas-phase absorber in the 200-250~nm range, which is covered by the \textsl{uvm2} filter \citep[S19,][]{Turner2016}. 
FeII absorption is generally predicted to increase with increasing irradiation -- as gauged by equilibrium temperature -- \textbf{but} the predicted NUV transit depth relative to the optical (600-700~nm) 
does not exceed 20\%, and peaks at $T_{\rm eq} \approx 2200$~K \citep{Lothringer2020}. 
If FeII absorption is responsible for the deep NUV transit of XO-3b, which has an estimated mass loss rate that is nearly 1000 times smaller than that of WASP-121b, then this work challenges the assumption that the strength of FeII absorption correlates with mass loss. 
This conclusion is supported by the theoretical work of \citet{Young2020}, which predicted significant FeII absorption using a CLOUDY model for the exosphere of HD~209458b, a well studied hot Jupiter orbiting an F-type star with a relatively small estimated mass-loss rate ($10^{10}$~g~s$^{-1}$).  
More work is needed, in both modeling individual star-planet systems and examining the transit signal under the relevant instrumental resolution, to say whether exospheric FeII can explain the NUV transit depths in this work.

\subsection{NUV transits as a probe of atmospheric aerosols}
\label{sec:Clouds}

Another explanation for a larger NUV radius relative to the optical could be the presence of aerosols with particle sizes $< 1~\mu{\rm m}$, which are highly efficient at attenuating NUV light. 
In general, clouds are expected to form more easily in environments with $T < 2000$~K, at which point aluminum oxide, iron, and silicate condensates can form \citep[e.g.,][]{Morley2012,Sing2016,Wakeford2017}. This motivates an examination of NUV transit depths as they depend on planet equilibrium temperature.

\begin{figure}[tp]
    \centering
    \includegraphics[width=.49\textwidth]{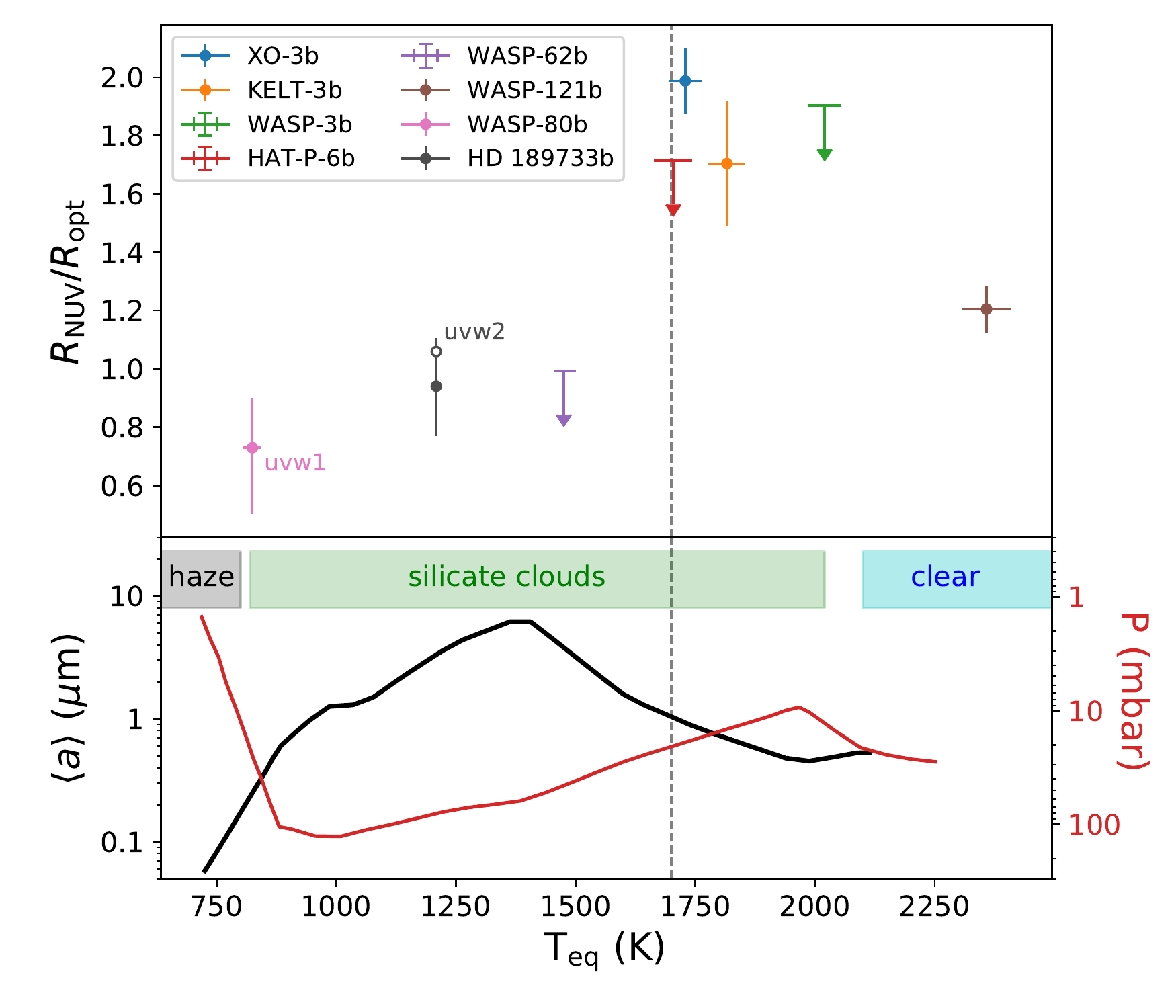}
    \caption{(Top) The ratio of the NUV and optical radius versus the planet equilibrium temperature for each of the hot Jupiters in our sample. Markers follow the same conventions as those in Figure~\ref{fig:insolation}. 
    \citep[Bottom, from Figure~2 of][]{Gao2020} The mean particle radius of aerosols probed by J/H band transmission spectroscopy (black curve). The atmospheric pressure at which NIR water features are obscured by aerosols is overplotted (red curve). The temperature zones mark which aerosol type accounts for $> 50\%$ of the total J/H band opacity, or where no aerosols are expected to form. The dashed vertical line marks the highest temperature at which the mean particle size is $\geq 1~\mu{\rm m}$.
    }
    \label{fig:Teq}
\end{figure}

We plot the ratio of the NUV and optical planet radius versus equilibrium temperature (Figure~\ref{fig:Teq}, top). 
We find that, for $T \gtrsim 1700$~K, the $R_{\rm NUV}/R_{\rm opt} > 1$ and can be as large as 2. For $T \lesssim 1700$~K, we have only two data points, exhibiting $R_{\rm NUV}/R_{\rm opt} \leq 1$. 
In a study of the XMM-OM transit of HD~189733b, which has $T_{\rm eq} = 1209 \pm 11$ \citep{Addison2019}, the NUV radius is equal the optical radius within $1\sigma$ uncertainty for the \textsl{uvm2} filter, and within $2\sigma$ uncertainty for the \textsl{uvw2} filter \citep{King2021}. This holds with the general observed trend.

Around 1700~K, a transition is also observed in the strength of the 1.4~$\mu{\rm m}$ water feature measured from hot Jupiters \citep{Fu2017}. 
One dimensional models of hot Jupiter atmospheres by \citet[][hereafter G20]{Gao2020} demonstrated that this could be explained by the onset of cloud formation, which can obscure the water signatures. 
The G20 results for cloud particles sizes and atmospheric pressures contributing to the NIR opacity near the water features of interest are plotted in Figure~\ref{fig:Teq} (bottom). In this scenario, water features become obscured as cloud particles -- mostly silicates -- form around 10~mbar. At lower temperatures, the cloud particles can reach sizes as large as 8~$\mu{\rm m}$, but sink to 100~mbar, allowing some water in the upper atmosphere to be visible. Below 1000~K, hydrocarbon hazes begin to form, dominating the opacity at the higher altitude 1-10~mbar atmosphere and obscuring the water features again.

The G20 models might also explain the observed trends in $R_{\rm NUV}/R_{\rm opt}$. The dashed vertical line in Figure~\ref{fig:Teq} ($T_{\rm eq} = 1700$~K) marks the highest temperature at which water obscuring cloud particles $> 1~\mu{\rm m}$ can form. This value happens to coincide with the peak $R_{\rm NUV}/R_{\rm opt}$ value.
Even though $1~\mu{\rm m}$ is larger than the grain size expected to maximize NUV attenuation, numerical models of cloud formation 
generally yield a high density cloud deck, full of large particles, with smaller particles extending into higher layers of the atmosphere. 
\citet{Powell2018} demonstrated that assuming a single mean particle size underestimated cloud opacities by a factor of 3-5 compared to using the full particle size distribution. Additionally, hotter atmospheres had cloud tops with smaller particles that extended to higher altitudes, leading to a steeper slope in the optical transmission spectrum \citep{Powell2018, Powell2019}. 
Thus at $T_{\rm eq} > 1700$~K, we expect a population of cloud particles with radii $< 1~\mu{\rm m}$ to contribute significantly to the NUV transit depth. 

\begin{figure}[tp]
    \centering
    \includegraphics[width=.49\textwidth]{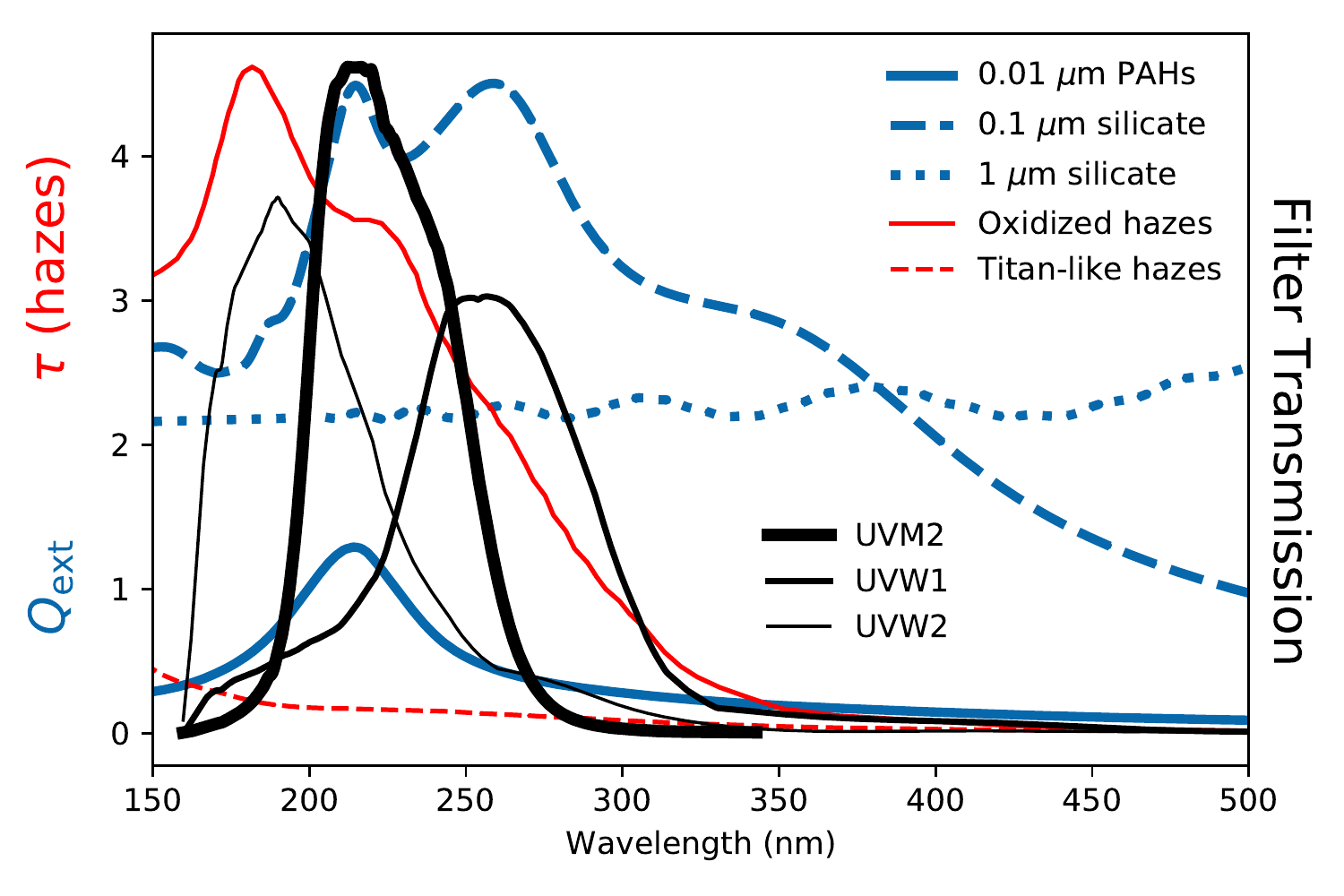}
    \caption{Transmission properties of three \textsl{Swift}-UVOT filters (black curves, almost identical to those of XMM-Newton OM), compared to the absorption profiles of exoplanet aerosol candidates. The extinction efficiency ($Q_{\rm ext}$, blue curves) is the extinction cross section area divided by the geometric cross section of the obscuring particles. The optical depth measured from thin films of lab produced hazes \citep{Gavilan2018} are plotted in red. The bluer filters (\textsl{uvw2} and \textsl{uvm2}) are more sensitive to extinction from PAHs (solid blue curve) and oxidized haxes (solid red curve) than the \textsl{uvw1} filter. All filters are more sensitive to small silicate particles (blue dashed curve) than large silicate particles (blue dotted curve), which remove all wavelengths of UV and visible light equally. 
    }
    \label{fig:Transmission}
\end{figure}

Cloud particles with $a \geq 1~\mu{m}$ are grey in opacity -- obscuring NUV and broad-band optical light equally (Figure~\ref{fig:Transmission}). Thus the NUV and optical transits of hot Jupiters in the $T_{\rm eq} \lesssim 1700$~K regime could probe similar atmospheric regions -- the atmosphere just above the cloud deck -- leading to NUV transit depths that are roughly equal to the optical transit depths. 
WASP-62b, which is near the temperature range where G20 predicts the largest cloud particles, exhibits pressure broadened NaI absorption features and has a high resolution transmission spectrum that can be fit without clouds \citep{Alam2020}. 
This suggests that any 
clouds that may be present on WASP-62b have sunk well below the pressure at which the atmosphere becomes optically thick.

At 800~K, where WASP-80b sits, the G20 model is at a transition point between a silicate cloud and haze dominated atmosphere. Because hazes are likely to be efficient NUV absorbers, one might expect the apparent NUV radius of the planet to be larger than the optical. However, the observed WASP-80b NUV transit is shallower than the optical transit. 
This could be because the WASP-80b observation was conducted with the \textsl{uvw1} filter, which is redder than the \textsl{uvm2} and extends 
significantly into the U-band. 
Figure~\ref{fig:Transmission} plots the extinction efficiency of policyclic aromatic hydrocarbons \citep[PAHs,][]{LD2001b}, which we treat as a proxy for large hydrocarbon molecules, along with the optical depths for samples of Titan-like and oxidized hazes produced in the lab \citep{Gavilan2018}. 
Because the \textsl{uvw1} filter misses the peak of these absorption profiles, its possible that the WASP-80b measurement happens to probe a relatively transparent window for typical atmospheric hazes.
Measuring transits of cooler hot Jupiters such as this one, in different NUV filters, could be important for examining the prevalence and composition of hazes in exoplanet atmospheres.

Despite the parallels between the NUV transit depth and the current state of cloud formation simulations, none of the transmission spectrum models presented for $T_{\rm eq} > 1800$~K hot Jupiter atmospheres suggest that NUV transit depths could be more than 10\% larger than the optical \citep{Powell2019}. 
However, lab experiments have demonstrated that hazes can form even in hot atmospheres around 1500~K \citep{He2020, Fleury2019}. 
To fully investigate the hypotheses that of aerosols could contribute to NUV transit depths, modeling the dual populations of cloud condensates and photochemically produced hazes is necessary. 
\citet{Helling2020} recently simulated these two physical mechanisms of aerosol production for the case of WASP-43b, which has $T_{\rm eq} \approx 1400$~K \citep{Esposito2017}. 
Using the photochemical models of \citet{Kawashima2018}, they predicted that hazes in WASP-43b could extend to atmospheric pressures that were 1-3 dex lower than the pressure levels occupied by cloud particles. However, their simulated hazes  contributed far less to the NUV atmospheric opacity compared to cloud particles. 
In addition to using a single particle size to estimate the haze opacity, \citet{Helling2020} used optical properties of Titan hazes, which are less absorbant in the NUV range than oxidized hazes \citep[][demonstrated in Figure~\ref{fig:Transmission}]{Gavilan2018}. 

HD~189733b is a hot Jupiter that is well known for having a steep optical slope in its transmission spectrum, which might be explained by atmospheric aerosols \citep{Sing2011, Pont2013, Barstow2020, SanchezLopez2020}.  
Using 3D circulation models of photochemically produced hazes, \citep{Steinrueck2021} demonstrated that enhanced vertical mixing, or optical properties of hazes that have stronger reddening properties than soot, are one of a few options for explaining the steep slope of optical transmission spectrum of HD~189733b. 
The error bars from broad-band NUV transit measurements obtained by \citet{King2021} are too large to place strong constraints on the aerosol type, and are consistent with the planet radius in the optical. However, it is worth noting that the best fit value for $R_p/R_*$ from the bluer \textsl{uvw2} filter (measured from 16 separate transit visits, resulting in a 5\% level precision) is larger than that obtained with \textsl{uvm2} (measured from a single transit, resulting in 15\% level precision). Such behavior might be expected from some species of aerosols or silicate grains smaller than 0.1~$\mu$m (Figure~\ref{fig:Transmission}).
Because of the broad diversity in planetary properties and radiation fields, detailed modeling of individual systems will be important to investigate whether aerosols can produce the NUV transit depths found in this work.

\subsection{Sources of systematic bias}
\label{sec:bias}

Heterogeneity on the stellar surface, such as starspots and faculae, are a source of systematic bias for transmission spectra of exoplanets. At  optical to infrared wavelengths, transit depths are only expected to be affected on the $\lesssim 5\%$ level for F stars such as those in this study \citep{Rackham2019}. 
At shorter wavelengths, the magnetic activity that produces dark starspots in the photosphere (optical) will appear as bright patches in the chromosphere (UV). 
If the exoplanet happens to pass in front of these active areas, the short wavelength transit will appear deeper than the true planet radius at that wavelength. However, the magnitude of the effect is inversely proportional to wavelength \citep{Llama2015}, and the 200-250~nm range probed by the \textsl{uvm2} filter is expected to be the least affected UV band.

Averaging over multiple transits also reduces the likelihood that a transit depth will be under- or overestimated. \citet{Llama2016} showed that, at Ly-$\alpha$ wavelengths (121~nm), combining the data from two transit visits was sufficient to recover the true planet radius 80\% of the time. Because the effect is less pronounced for longer wavelengths, we expect our 2-3 visit lightcurves to yield a reliable transit depth. 
Assessing the effect of starspots and stellar variability on our targets would require additional transit visits and stellar activity monitoring. % with a more powerful telescope than \textsl{Swift}.

Finally, we note that the object in our study with the highest $R_{\rm opt}/R_{\rm NUV}$ value, XO-3b, is also the planet that is most subject to heating by tidal dissipation. The fact that the observed NUV transit occurs earlier than predicted optical transits hints that the NUV absorption might arise from other gas in the system, such as a compressed bow-shock formed as the planet's atmosphere interacts with the a stellar wind \citep[e.g.,][]{Llama2013}. Thus the uniqueness of this result might be a symptom of the uniqueness of this planetary system.

\section{Conclusion} \label{sec:conclude}

Selecting for NUV brightness of the host star, we used \textsl{Swift}-UVOT to search for NUV transit signals of five hot Jupiters not yet reported in the literature.
This gave us a relatively unbiased sample with which to explore the properties of NUV absorption in hot Jupiters and its relation to global properties such as insolation, mass-loss, and equilibrium temperature. 
We demonstrate the feasibility of our analysis pipeline by reproducing the transit measurement of WASP~121b \citep{Salz2019}, showing that careful aperture photometry with standard \textsl{Swift}-UVOT analysis tools, combined with reference lightcurves from nearby field stars, can measure a $\sim 1\%$ level transit to $> 3\sigma$ significance. 
We applied these techniques to our sample of transit lightcurves, obtaining one positive $3\sigma$ transit detection (XO-3b), one marginal $1-2\sigma$ detection (KELT-3b), and three upper limits (WASP-3b, HAT-P-6b, and WASP-62b). 
We attribute the null detections of WASP-3b and WASP-62b to sparsely sampled lightcurves. 

We used the simultaneous \textsl{Swift}-XRT data to measure the X-ray luminosity of the exoplanet host stars. In four of the cases, no X-ray point source was found and we are only able to place upper limits. In the case of WASP-62, we detect a point source at the $2.5\sigma$ level. From this detection we estimate an unabsorbed 0.2-2.4~keV flux of $1.7 (\pm 0.8) \times 10^{-14}$~erg~cm$^{-2}$~s$^{-1}$, which implies $L_{\rm X}(0.2-2.4~{\rm keV} = 6 (\pm 3) \times 10^{28}$~erg~s$^{-1}$. 
However, we extracted a higher quality spectrum from the publicly available XMM-Newton data available for WASP-62b. Fitting the spectrum gives us a more precise value, $L_{\rm X}(0.2-2.4~{\rm keV}) = 9.3_{-1.3}^{+1.9}$~erg~s$^{-1}$.
Following the procedures of \citet{King2018}, we used the 0.2-2.4~keV insolation on the exoplanets to estimate the EUV irradiation incident on each planet and the corresponding energy limited mass-loss rate.

We examined the apparent NUV radius of each planet relative to the apparent optical radius ($R_{\rm NUV}/R_{\rm opt}$), which controls for uncertainty in the host star stellar radius. We find no apparent correlation of $R_{\rm NUV}/R_{\rm opt}$ with X-ray insolation. In particular, XO-3b, which has the smallest estimated mass-loss rate in this sample also has the largest measured NUV transit depth. 
This finding contradicts the hypothesis that NUV transit depth is positively correlated with photoevaporative mass loss.

We also examined $R_{\rm NUV}/R_{\rm opt}$ as a function of exoplanet equilibrium temperature. We find a tentative trend: $R_{\rm NUV}/R_{\rm opt} > 1$ for $T_{\rm eq} \gtrsim 1700$~K. After that, we see a decrease to $R_{\rm NUV}/R_{\rm opt} \leq 1$ for $T_{\rm eq} \lesssim 1700$~K. This trend appears analogous to that found for water absorption strengths, which is highly sensitive to obscuration by aerosols \citep{Fu2017, Gao2020}. 
We interpret our results under the paradigm put forward by \citet{Gao2020}. Exoplanets with $T_{\rm eq} \approx 1700-2100$~K have high altitude cloud decks dominated by $\leq 1~\mu{\rm m}$ sized silicate particles that are efficient at attenuating NUV light. Below 1700~K, the cloud particles grow so large that they obscure both NUV and optical light equally and sink lower in the atmosphere. In these temperature ranges, we expect that the NUV and optical light are probing the exoplanet atmosphere at nearly identical scale heights. 

Finally, we demonstrate that the NUV filter wheels available on \textsl{Swift}-UVOT might be sensitive to different species of aerosols. 
The particularly low NUV transit depth observed for WASP-80b might be explained by the fact that the NUV dataset was obtained with the {\sl uvw1} filter, which suffers significant contamination by the U-band and is red-ward of the strongest FeII and haze absorption features. The recent XMM-Newton OM observations of HD~189733b \citep{King2021} in two filters, \textsl{uvw2} and \textsl{uvm2}, are also consistent with the tentative trends in $T_{\rm eq}$ described above. A more detailed examination of aerosol transmission spectroscopy models covering the wavelength range of the \textsl{Swift}-UVOT filters will be the subject of a future work.

This work demonstrates the expanded capabilities of \textsl{Swift}-UVOT to observe planetary transits in the NUV. Obtaining planetary radii at short wavelengths such as these can provide insight into crucial processes of atmospheric evolution, such as photoevaporative mass-loss and cloud formation. The complementary X-ray data provided by \textsl{Swift} is also important for quantifying the mass-loss rates and understanding the impact of high energy irradiation on planets.

\acknowledgments

 Thank you to the anonymous reviewers for their questions that greatly helped improve the clarity of the paper. We acknowledge the use of public data from the \textsl{Swift} data archive. We thank Sam LaPorte from the \textsl{Swift} scheduling team, Michael Siegel, and the \textsl{Swift}-UVOT team for their support. We would also like to thank Evgenya Shkolnik for particularly helpful conversation about stellar activity in the UV. This research has made use of the SVO Filter Profile Service\footnote{\href{http://svo2.cab.inta-csic.es/theory/fps/}{http://svo2.cab.inta-csic.es/theory/fps/}} supported from the Spanish MINECO through grant AYA2017-84089. Partial support for this work was provided by grant number AR6-17006B awarded by the Chandra X-ray Center (CXC), which is operated by the Smithsonian Astrophysical Observatory for NASA under contract NAS8-03060.

\vspace{5mm}
\facilities{The Neil Gehrels Swift Observatory}

\software{\href{https://heasarc.gsfc.nasa.gov/docs/software/heasoft/}{HEAsoft}, $\;$ \href{www.astropy.org}{Astropy} \citep{Astropy2}, $\;$ \href{https://www.cfa.harvard.edu/~lkreidberg/batman/}{{BATMAN}} \citep{Kreidberg2015}, $\;$ 
\href{https://emcee.readthedocs.io/en/stable/}{{emcee}} \citep{emcee}, $\;$  
\href{https://exoctk.stsci.edu/}{{ExoCTK}}, $\;$ 
\href{https://github.com/elizakempton/Exo_Transmit}{{Exo-Transmit}} \citep{Kempton2017}}

\appendix

\section{Instrumental Correction with Reference Light Curves}

As described in Section~\ref{sec:analysis}, each target star light curve was normalized by a reference light curve produced by averaging the flux measurements of a collection of field stars that were visible in all of the event files. This method was implemented in order to correct for any systematic bias arising from the instrument, such as gain fluctuations. The number of reference field stars available for the correction process, for each transit target, are listed in the eighth column of Table~\ref{tab:vislist}.

\begin{figure*}[h]
    \centering
    \includegraphics[width=.98\textwidth]{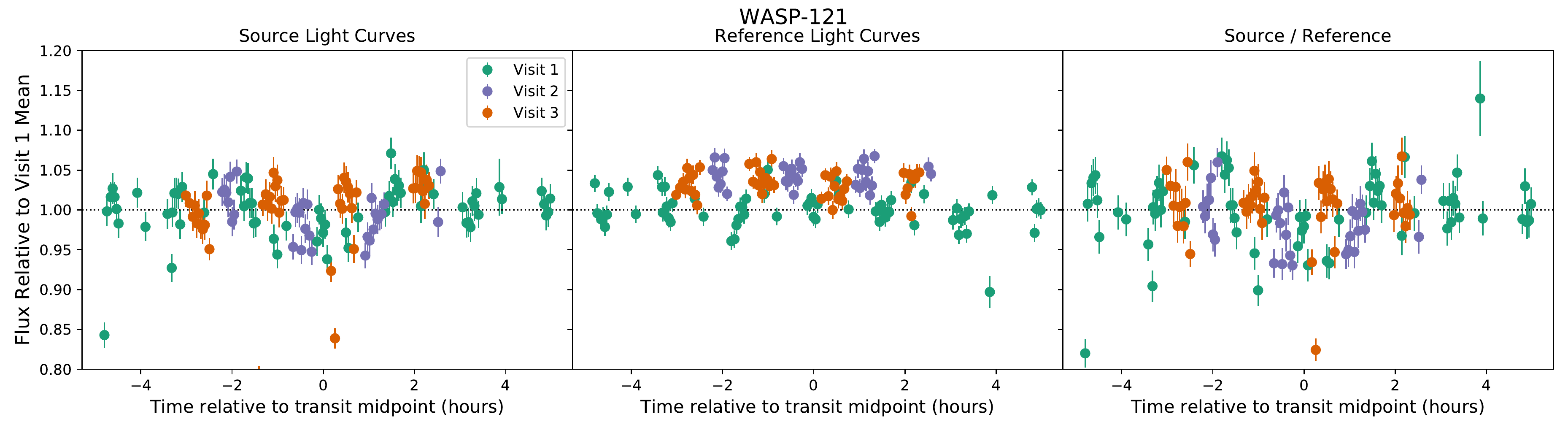}
    \caption{(Left) Light curves from each visit of the target source, WASP-3. (Middle) Light curve obtained from averaging the flux measurements of the reference field stars within each time bin, for WASP-3. (Right) Corrected light curve for WASP-3, obtained by normalizing the source light curve by the reference light curve. In each panel, all light curves are normalized by the mean flux value of Visit 1, in order to illustrate relative changes in the extracted flux values across different visits. 
    }
    \label{fig:WASP121-corr}
\end{figure*}

\begin{figure*}[h]
    \centering
    \includegraphics[width=.98\textwidth]{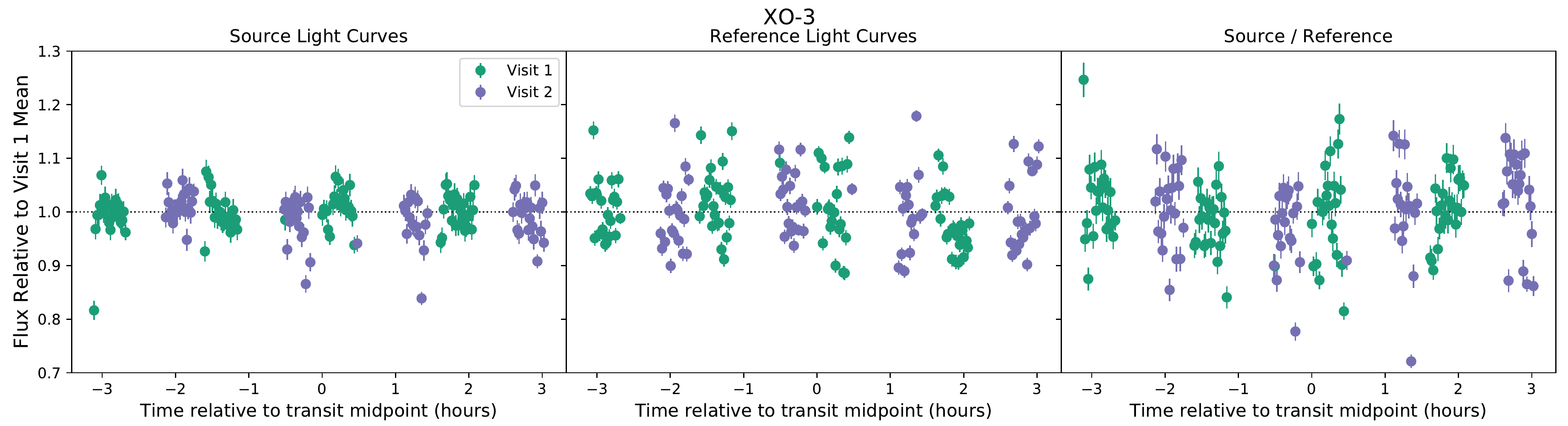}
    \caption{Same as Figure~\ref{fig:WASP121-corr} but for XO-3. 
    }
    \label{fig:XO3-corr}
\end{figure*}

\begin{figure*}[h]
    \centering
    \includegraphics[width=.98\textwidth]{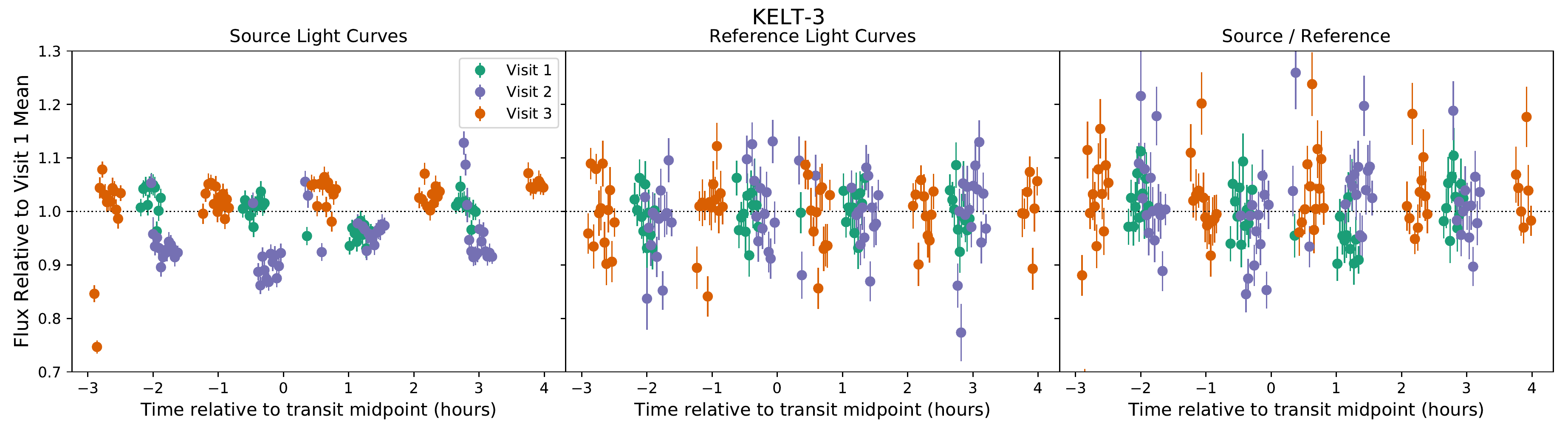}
    \caption{Same as Figure~\ref{fig:WASP121-corr} but for KELT-3. 
    }
    \label{fig:KELT3-corr}
\end{figure*}

\begin{figure*}[h]
    \centering
    \includegraphics[width=.98\textwidth]{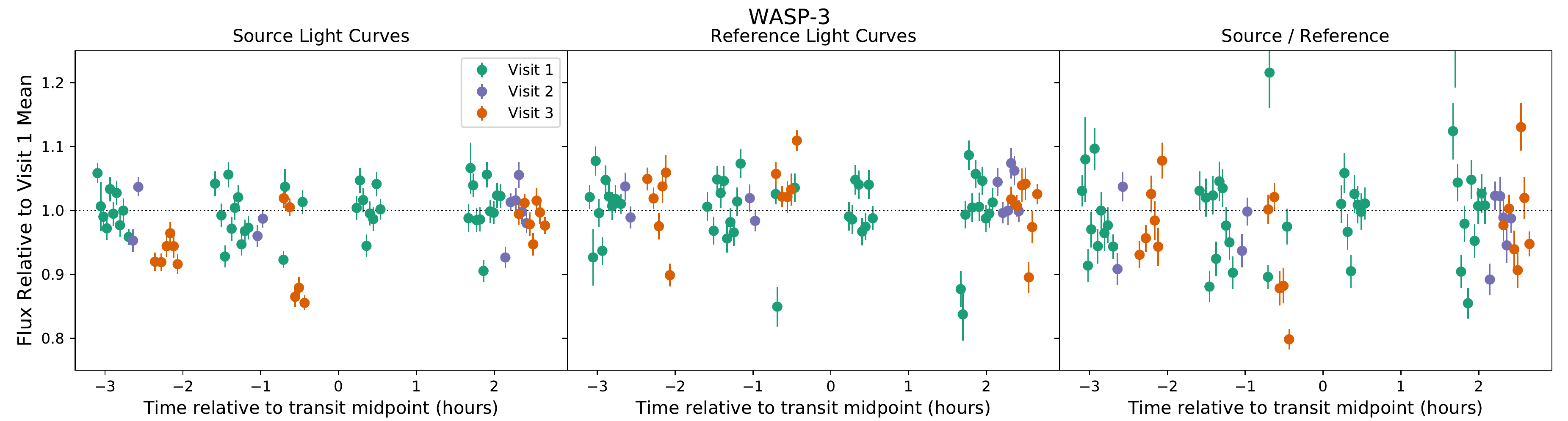}
    \caption{Same as Figure~\ref{fig:WASP121-corr} but for WASP-3. 
    }
    \label{fig:WASP3-corr}
\end{figure*}

\begin{figure*}[h]
    \centering
    \includegraphics[width=.98\textwidth]{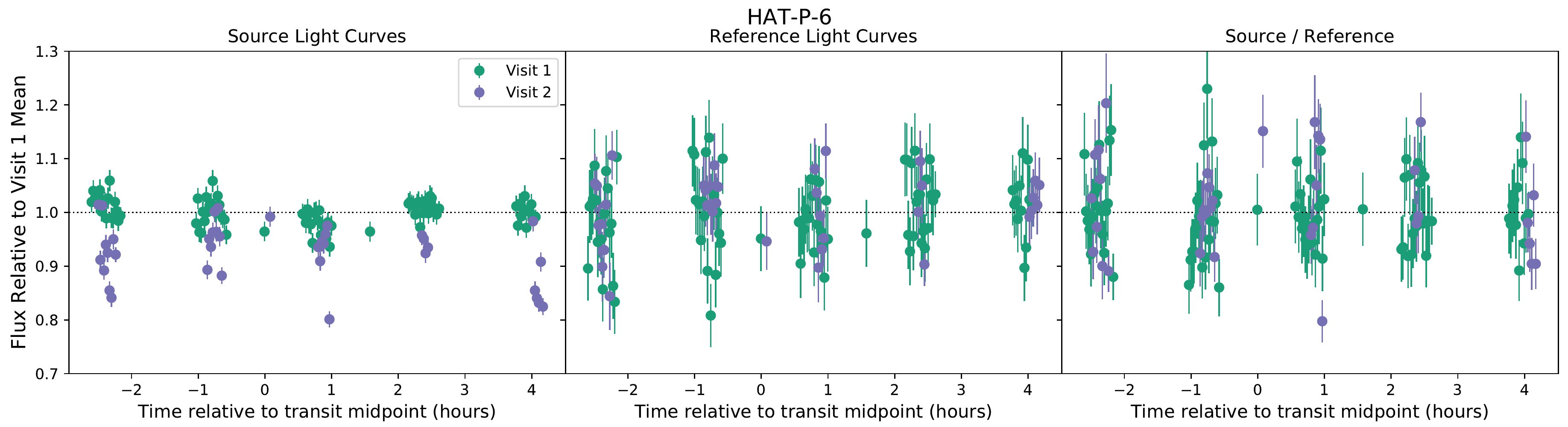}
    \caption{Same as Figure~\ref{fig:WASP121-corr} but for HAT-P-6. 
    }
    \label{fig:HATP6-corr}
\end{figure*}

\begin{figure*}[h]
    \centering
    \includegraphics[width=.98\textwidth]{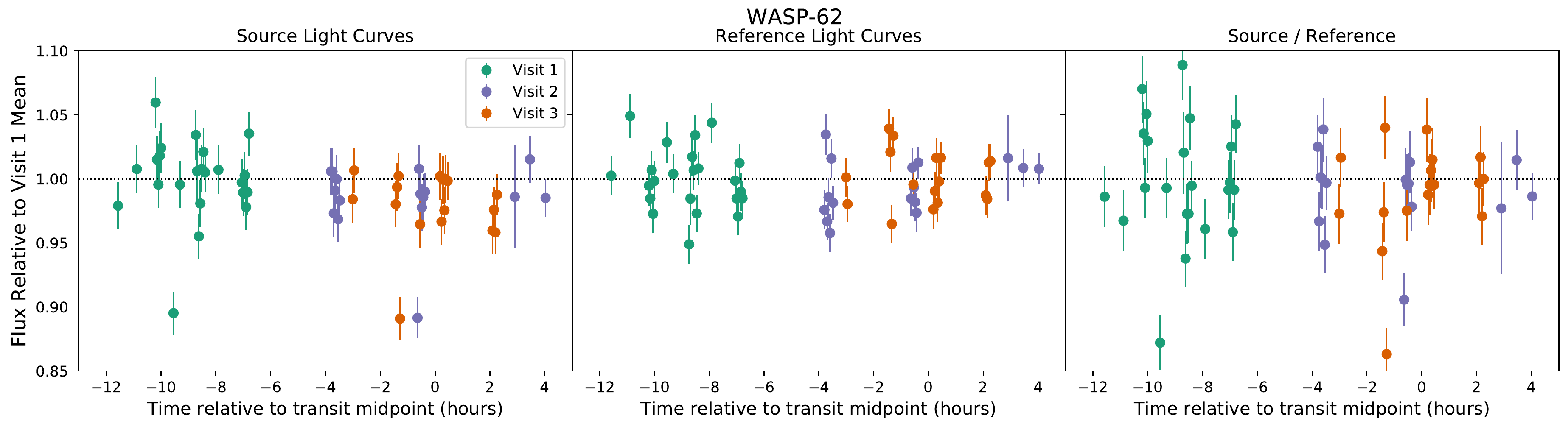}
    \caption{Same as Figure~\ref{fig:WASP121-corr} but for WASP-62. 
    }
    \label{fig:WASP62-corr}
\end{figure*}

Figures~\ref{fig:WASP121-corr}-\ref{fig:WASP62-corr} illustrate the results of the instrument correction process. Left panels show the original target star flux light curve. Middle panels show the reference light curve obtained by averaging the flux values of the reference stars within each time bin. The right panels show the final light curves that result from the correction process of normalizing the source by the reference values. For visual purposes, the y-axis (flux) values are normalized by the mean flux of the "Visit 1" light curves within each panel, separately. This illustrates absolute changes in the measured flux across different transit visits.

The process described above appears to correct for absolute brightness of the target stars across different visits. For example, the target light curve from Visit~2 of KELT-3 (purple markers in Figure~\ref{fig:KELT3-corr}) appears 5-10\% dimmer in absolute flux compared to Visit~1. However, after normalizing the source light curve by the reference light curve, the absolute brightness of KELT-3 in Visit~2 appears more centered on the average value of Visit~1.

Because the reference field stars are usually dimmer than the target star, and because a small number of reference stars are available, the scatter and error bar size on the reference light curves are generally larger than that found in the target source light curves. As a result, the correction process increases the error bar size and scatter (i.e., sacrificing precision for accuracy) in the final light curves used for the transit searches.

\bibliography{references.bib}

\end{document}